\documentclass[11pt,a4paper]{article}
\usepackage{jcappub}
\usepackage{amsmath}
\usepackage{amssymb}
\usepackage{dcolumn}
\usepackage{bm}
\usepackage{color}
\usepackage{epsfig}
\usepackage{amsfonts}

\newcommand{\ud}{\mathrm{d}}

\title{Phase-Space analysis of Teleparallel Dark Energy}
\author[a]{Chen Xu,}
\author[b,c]{Emmanuel N. Saridakis}
\author[d]{and Genly Leon}
\affiliation[a]{College of Mathematics and Physics, Chongqing
University of Posts and \\ Telecommunications, Chongqing 400065,
China} \affiliation[b]{Physics Division, National Technical
University of Athens, 15780 Zografou Campus,  Athens, Greece}
\affiliation[c]{CASPER, Physics Department, Baylor University,
Waco, TX  76798-7310, USA} \affiliation[d]{Department of
Mathematics, Universidad Central de Las Villas, Santa Clara
\enskip CP 54830, Cuba} \emailAdd{xuc1990@126.com}
\emailAdd{Emmanuel$_-$Saridakis@baylor.edu}
\emailAdd{genly@uclv.edu.cu}
\abstract{We perform a detailed
dynamical analysis of the teleparallel dark energy scenario, which
is based on the teleparallel equivalent of General Relativity, in
which one adds a canonical scalar field, allowing also for a
nonminimal coupling with gravity. We find that the universe can
result in the quintessence-like, dark-energy-dominated solution,
or to the stiff dark-energy late-time attractor, similarly to
standard quintessence. However, teleparallel dark energy possesses
an additional late-time solution, in which dark energy behaves
like a cosmological constant, independently of the specific values
of the model parameters. Finally, during the evolution the dark
energy equation-of-state parameter can be either above or below
$-1$, offering a good description for its observed dynamical
behavior and its stabilization close to the cosmological-constant
value.}
\arxivnumber{1202.3781}
\begin{document}
\maketitle

\section{Introduction}

According to observations, the observable universe experiences an
accelerating expansion at late times \cite{Riess:1998cb,Perlmutter:1998np,Bennett:2003bz}. There are two major
ways to
explain such a behavior, apart from the simple
consideration of a cosmological constant. The first is to modify the
gravitational sector itself (see \cite{Nojiri:2006ri} for a review and
references therein), obtaining a modified cosmological dynamics. The
other approach is to modify the content of the universe introducing the
dark energy with negative pressure, which can be based on a canonical
scalar field
(quintessence)
\cite{Ratra:1987rm,Wetterich:1987fm,Zlatev:1998tr,Boisseau:2000pr,
Guo:2006ab,Dutta:2009yb,Sahni:1998at,Uzan:1999ch,Faraoni:2000wk,
Gong:2002dk,Elizalde:2004mq,Faraoni:2004dn,Setare:2008pc}, a phantom field
\cite{Caldwell2002a,Caldwell:2003vq,Nojiri:2003vn,Onemli:2004mb,Saridakis:2009pj,Dutta:2009dr}, or on the combination of both these fields in a unified
scenario called quintom
\cite{Feng:2004ad,Guo:2004fq,Feng:2004ff,Zhao:2006mp,Lazkoz:2006pa,
Lazkoz:2007mx,Saridakis:2009jq,Setare:2008si,Cai:2009zp} (see the reviews
\cite{Copeland:2006wr} and \cite{Leon:2009ce}).

``Teleparallel'' dark energy is a recently proposed scenario that
tries to incorporate the dark energy sector \cite{Geng:2011aj,Geng:2011ka}.
It lies on the framework of the
``teleparallel'' equivalent of General Relativity (TEGR), that is based on
torsion instead of curvature formulation
\cite{Unzicker:2005in,Hayashi:1979qx}, in which
one adds a canonical scalar field, and the dark energy sector is attributed
to this field. When the field is minimally-coupled to gravity
the scenario is completely equivalent to standard quintessence
\cite{Ratra:1987rm,Wetterich:1987fm,Zlatev:1998tr},
both at the background and perturbation levels.
However, in the case where one allows for a nonminimal coupling between
the field and the torsion scalar which is the only suitable gravitational
scalar in TEGR, the resulting scenario presents a richer structure,
exhibiting
quintessence-like or phantom-like behavior, or experiencing the
phantom-divide crossing \cite{Geng:2011aj}.

The interesting cosmological behavior of teleparallel dark energy makes it
necessary to perform a phase-space and stability
analysis, examining in a systematic way the possible cosmological
behaviors, focusing on the late-time stable solutions. In the present
work we perform such an approach, which allows us to bypass the high
non-linearities of the cosmological equations,
which prevent any complete analytical treatment, obtaining a
(qualitative) description of the global dynamics of these models, that is
independent of the initial conditions and the specific evolution of the
universe. Furthermore, in these asymptotic solutions we calculate various
observable quantities, such as the deceleration parameter, the
effective (total) equation-of-state parameter, and the various
density parameters.

The plan of the work is the following: In section \ref{TEGRquint}
we briefly review the scenario of teleparallel dark energy and in section
\ref{phasespace} we perform a detailed phase-space analysis. In section
\ref{implications} we discuss the cosmological implications and the
physical behavior of the scenario. Finally, section \ref{Conclusions} is
devoted to the summary of the results.

\section{Teleparallel Dark Energy}
\label{TEGRquint}

In this section we review teleparallel dark energy. Such a scenario is
based on the ``teleparallel'' equivalent of General Relativity
(TEGR)\cite{Unzicker:2005in,Hayashi:1979qx}, in which instead of using
the torsionless
Levi-Civita connection one uses the curvatureless Weitzenb{\"o}ck one. The
dynamical objects are the four linearly independent, \emph{parallel} vector
fields called vierbeins, and the advantage of this framework is that the
torsion tensor is formed solely from products of first derivatives of the
tetrad. In particular, the vierbein field ${\mathbf{e}_A(x^\mu)}$ forms an
orthonormal basis for the tangent space at each point $x^\mu$, that is
$\mathbf{e}
_A\cdot
\mathbf{e}_B=\eta_{AB}$, where $\eta_{AB}=diag (1,-1,-1,-1)$, and
furthermore the vector $\mathbf{e}_A$ can be analyzed with the use of
its components $e_A^\mu$ in a coordinate basis, that is
$\mathbf{e}_A=e^\mu_A\partial_\mu $
\footnote{In this manuscript we follow the notation of
\cite{Chen:2010va,Dent:2011zz,Cai:2011tc}, that is Greek indices $\mu, \nu,$... and capital Latin
indices $A, B, $... run over all coordinate and tangent
space-time 0, 1, 2, 3, while lower case Latin indices (from the middle of
the alphabet) $i, j,...$ and lower case Latin indices (from the beginning
of the alphabet) $a,b, $... run over spatial and tangent
space coordinates 1, 2, 3, respectively. Finally,
we use the metric signature $(+,-,-,-)$.}.
The metric tensor is obtained from the
dual vierbein
as
\begin{equation}  \label{metrdef}
g_{\mu\nu}(x)=\eta_{AB}\, e^A_\mu (x)\, e^B_\nu (x).
\end{equation}
Furthermore, the  torsion tensor of the Weitzenb%
\"{o}ck connection $\overset{\mathbf{w}}{\Gamma}^\lambda_{
\nu\mu}$ \cite{Weitzenb23} reads
\begin{equation}  \label{torsion2}
{T}^\lambda_{\:\mu\nu}=\overset{\mathbf{w}}{\Gamma}^\lambda_{
\nu\mu}-%
\overset{\mathbf{w}}{\Gamma}^\lambda_{\mu\nu}
=e^\lambda_A\:(\partial_\mu
e^A_\nu-\partial_\nu e^A_\mu).
\end{equation}

In such a framework all the information
concerning the
gravitational field is included in the torsion tensor
${T}^\lambda_{\:\mu\nu} $. The corresponding ``teleparallel Lagrangian''
can be constructed from this torsion tensor under the assumptions of
invariance under general coordinate transformations, global Lorentz
transformations, and the parity operation, along with requiring the
Lagrangian density to be second order in the torsion tensor
\cite{Hayashi:1979qx}. In particular, it is proved to be just the torsion scalar
$T$, namely
\cite{Unzicker:2005in,Hayashi:1979qx,Maluf:1994ji,Arcos:2005ec}:
{\small{
\begin{equation}  \label{telelag}
\mathcal{L}=T=\frac{1}{4}T^{\rho
\mu \nu }T_{\rho \mu \nu }+\frac{1}{2}T^{\rho \mu \nu
}T_{\nu \mu \rho }-T_{\rho \mu }^{\ \ \rho }T_{\ \ \ \nu }^{\nu
\mu }.
\end{equation}}}
Therefore, the simplest action in a universe governed by teleparallel
gravity is
\begin{eqnarray}  \label{action}
I =\int d^4x e
\left[\frac{T}{2\kappa^2}+\mathcal{L}_m\right],
\end{eqnarray}
where $e = \text{det}(e_{\mu}^A) = \sqrt{-g}$ (one could also include a
cosmological constant). If we vary it with respect to the vierbein
fields we obtain the equations of
motion, which are exactly the same as those of General Relativity for every
geometry choice, and that is why the theory is called ``teleparallel
equivalent to General Relativity''.

At this stage one has two ways of generalizing the action (\ref{action}),
inspired by the corresponding procedures of standard General Relativity.
The first is to replace $T$ by an arbitrary function $f(T)$
\cite{Bengochea:2008gz,Linder:2010py,Myrzakulov:2010vz,Wu:2010av,
Bamba:2010iw,Zheng:2010am,Bamba:2010wb,Wang:2011xf,Yerzhanov:2010vu,
Yang:2010ji,Wu:2010mn,Bengochea:2010sg,Wu:2010xk,Li:2010cg,Zhang:2011qp,
Deliduman:2011ga,Chattopadhyay:2011fp,Sharif:2011bi,Li:2011rn,Wei:2011jw,
Ferraro:2011zb,Miao:2011ki,Boehmer:2011gw,Wei:2011mq,Capozziello:2011hj,
Wu:2011xa,Daouda:2011rt,Bamba:2011pz,Wu:2011kh,Gonzalez:2011dr,
Ferraro:2011ks,Boehmer:2011si,Karami:2011nj,Wei:2011aa,Atazadeh:2011aa,
Farajollahi:2011af,Jamil:2011mc,Karami:2012fu,Yang:2012hu,Daouda:2012nj,
Iorio:2012cm,Chen:2010va,Dent:2011zz,Cai:2011tc}, similarly to $f(R)$
extensions of GR,
and obtain
new interesting terms in the field equations. The second, on which we focus
in this work, is to add a canonical scalar field in (\ref{action}),
similarly to GR quintessence, allowing for a nonminimal coupling with
gravity. Then the dark energy sector will be attributed to this field and
therefore the corresponding scenario is called ``teleparallel dark energy''
\cite{Geng:2011aj}.
In particular, the action will simply read:
\begin{equation}
S=\int\ud^{4}x e\Bigg[\frac{T}{2\kappa^{2}}
+ \frac{1}{2} \Big(\partial_{\mu}\phi\partial^{\mu}\phi+\xi
T\phi^{2}\Big) - V(\phi)+\mathcal{L}_m\Bigg]. \label{action2}
\end{equation}
Here we mention that since in TEGR, that is in the torsion formulation of
GR, the only scalar is the torsion one,  the
nonminimal coupling will be between this and the scalar field (similarly to
standard nonminimal quintessence where the scalar field couples to the
Ricci scalar).

Variation of action (\ref{action2}) with respect to the vierbein fields
yields the equation of motion
\begin{eqnarray}\label{eom2}
\left(\frac{2}{\kappa^2}+2 \xi
\phi^2 \right)\left[e^{-1}\partial_{\mu}(ee_A^{\rho}S_{\rho}{}^{\mu\nu} )
-e_{A}^{\lambda}T^{\rho}{}_{\mu\lambda}S_{\rho}{}^{\nu\mu}
-\frac{1}{4}e_{A}^{\nu
}T\right]\nonumber\\
-
e_{A}^{\nu}\left[\frac{1}{2}
\partial_\mu\phi\partial^\mu\phi-V(\phi)\right]+
  e_A^\mu \partial^\nu\phi\partial_\mu\phi\ \ \nonumber\\
+ 4\xi e_A^{\rho}S_{\rho}{}^{\mu\nu}\phi
\left(\partial_\mu\phi\right)
=e_{A}^{\rho}\overset {\mathbf{em}}T_{\rho}{}^{\nu}.~~~~~~~~~~~~
\end{eqnarray}
where
$\overset{\mathbf{em}}{T%
}_{\rho}{}^{\nu}$ stands for the usual
energy-momentum tensor. Thus, imposing a flat
Friedmann-Robertson-Walker (FRW) background metric
\begin{equation}
\label{FRWmetric}
ds^2= dt^2-a^2(t)\,\delta_{ij} dx^i dx^j,
\end{equation}
where $t$ is the cosmic time, $x^i$ are the comoving spatial
coordinates and $a(t)$ is the scale factor,
that is for a vierbein choice of the form
\begin{equation}  \label{FRWvierbeins}
e_{\mu}^A=\mathrm{diag}(1,a,a,a),
\end{equation}
we acquire the corresponding Friedmann equations:
\begin{eqnarray}
\label{FR1}
&&
H^{2}=\frac{\kappa^2}{3}\Big(\rho_{\phi}+\rho_{m}\Big),
\\
\label{FR2}
&&
\dot{H}=-\frac{\kappa^2}{2}\Big(\rho_{\phi}+p_{\phi}+\rho_{m}+p_{m}
\Big),~~~~
\end{eqnarray}
where $H=\dot{a}/a$ is the Hubble parameter,
with a dot denoting
$t$-differentiation.
 In these expressions,
 $\rho_m$ and $p_m$ are the matter energy density and
 pressure, respectively,  following the standard evolution equation
$\dot{\rho}_m+3H(1+w_m)\rho_m=0$, with $w_m= p_m/\rho_m$
the matter equation-of-state parameter. Furthermore, we have introduced
the effective energy density and pressure of scalar field
\begin{eqnarray}
\label{telerho}
 &&\rho_{\phi}=  \frac{1}{2}\dot{\phi}^{2} + V(\phi)
-  3\xi H^{2}\phi^{2},\\
 &&p_{\phi}=  \frac{1}{2}\dot{\phi}^{2} - V(\phi) +   4 \xi
H \phi\dot{\phi}
 + \xi\left(3H^2+2\dot{H}\right)\phi^2.\ \ \ \ \
 \label{telep}
\end{eqnarray}
Moreover, variation of the action with respect to the scalar field
provides its evolution equation, namely:
\begin{equation}
\ddot{\phi}+3H\dot{\phi}+6\xi H^2\phi+   V'(\phi)=0.
\label{fieldevol2}
\end{equation}
Note that in the above relations we have made use of the useful
relation
\begin{equation}
T=-6H^2,
\end{equation}
 which straightforwardly arises from
the calculation of
(\ref{telelag})
for the flat FRW geometry.

In the present cosmological paradigm, similarly to the standard
quintessence, dark
energy is attributed to the scalar field, and thus its equation-of-state
parameter reads:
\begin{equation}\label{EoS}
w_{DE}\equiv w_\phi=\frac{p_\phi}{\rho_\phi}.
\end{equation}
 Finally, one can see that the scalar field evolution (\ref{fieldevol2})
 leads to the standard relation
\begin{equation}\dot{\rho}_\phi+3H(1+w_\phi)\rho_\phi=0.
\end{equation}

Teleparallel dark energy exhibits very interesting cosmological
behavior. In the minimally-coupled case the cosmological equations coincide
with those of the standard quintessence, both at the background and
perturbation levels. However, in the nonminimal
case one can obtain a dark energy sector being
quintessence-like, phantom-like, or experiencing the phantom-divide
crossing during evolution, a behavior that is much richer comparing to
General Relativity (GR) with a scalar field \cite{Geng:2011aj,Geng:2011ka}.
Therefore, it is
  interesting to perform a phase-space analysis, that is
to investigate late-time solutions that are independent from the
initial conditions and the specific universe evolution. This is performed
in the next section.

\section{Phase-space analysis}
\label{phasespace}

In order to perform the phase-space and stability analysis of the scenario
at hand, we have to transform the aforementioned dynamical system into its
autonomous form $\label{eomscol} \textbf{X}'=\textbf{f(X)}$
\cite{Copeland:1997et,Ferreira:1997au,Gong:2006sp,Chen:2008pz,Chen:2008ft},
where $\textbf{X}$ is the column vector constituted by suitable
auxiliary variables, \textbf{f(X)} the corresponding  column
vector of the autonomous equations, and prime denotes derivative
with respect to $M=\ln a$. Then we extract its critical points
$\bf{X_c}$  satisfying $\bf{X}'=0$, and in order to determine
the stability properties of these critical points, we expand around
$\bf{X_c}$, setting
$\bf{X}=\bf{X_c}+\bf{U}$, with $\textbf{U}$ the perturbations of
the variables considered as a column vector. Thus, for each
critical point we expand the equations for the perturbations up to
the first order as: $\label{perturbation} \textbf{U}'={\bf{Q}}\cdot
\textbf{U}$, where the matrix ${\bf {Q}}$ contains the coefficients of the
perturbation equations. Finally, for each critical point, the eigenvalues
of ${\bf {Q}}$ determine its type and stability.

In the scenario at hand, we introduce the auxiliary variables:
\begin{eqnarray}
&&x=\frac{\kappa\dot{\phi}}{\sqrt{6}H}\nonumber\\
&&y=\frac{\kappa\sqrt{V(\phi)}}{\sqrt{3}H}\nonumber\\
&&z=\sqrt{|\xi|}\kappa\phi. \label{auxiliary}
\end{eqnarray}
Using these variables the Friedmann equation \eqref{FR1} becomes
\begin{equation}\label{Fr1norm}
x^2+y^2-z^2 \text{sgn}(\xi )+\frac{\kappa^2 \rho_m}{3 H^2}=1.
\end{equation}
This constraint allows for expressing $\rho_m$ as a function of the
auxiliary variables (\ref{auxiliary}). Therefore, using (\ref{Fr1norm})
and (\ref{telerho}) we can write the density parameters as:
\begin{eqnarray}
&&\Omega_m\equiv\frac{\kappa^{2}\rho_m}{3H^{2}}=1-x^2-y^2+z^2 \text{sgn}(\xi )\nonumber\\
 &&\Omega_{DE}\equiv\frac{\kappa^{2}\rho_{\phi}}{3H^{2}}=x^2+y^2-z^2 \text{sgn}(\xi ),
 \label{Omegas}
\end{eqnarray}
while for the dark-energy equation-of-state parameter (\ref{EoS}) we
obtain:

\begin{equation}\label{wdephase}
w_{DE}=\frac{x^2-y^2+4\sqrt{\frac{2}{3}}zx\sqrt{|\xi|}\text{sgn}
(\xi)-z^2w_m
\text{sgn}(\xi)\left[1-x^2-y^2+z^2\text{sgn}(\xi)\right]}{\left[1+z^2\text{
sgn}(\xi)
\right]\left[x^2+y^2-z^2\text{sgn}(\xi)\right] }.
\end{equation}

In this scenario $w_{DE}$ can
be quintessence-like, phantom-like, or experience the phantom divide
crossing during the evolution. Without loss of generality, in the following
we restrict the analysis in the dust matter case, that is we assume that
$w_m=0$.

It is convenient to introduce two additional quantities with great physical
significance, namely the ``total'' equation-of-state parameter:
{\small{
\begin{equation}\label{wtot}
w_{tot}\equiv\frac{p_\phi}{\rho_\phi+\rho_m}=w_{DE}\Omega_{DE}
=\frac{x^2-y^2+4\sqrt{\frac{2}{ 3}}zx\sqrt{|\xi|}\text{sgn}
(\xi)}{1+z^2\text{
sgn}(\xi)},
\end{equation}}}
and the deceleration parameter $q$:
\begin{eqnarray}
q&\equiv&-1-\frac{\dot{H}}{H^2}=\frac{1}{2}+\frac{3}{2}w_{tot}\nonumber\\
&=&
\frac{1+3(x^2-y^2)+\left[z+4\sqrt{6}x\sqrt{|\xi|}\right]z\text {
sgn}(\xi)}{2\left[
1+z^2\text {
sgn}(\xi)
\right]}.
\label{decc}
\end{eqnarray}

Finally, concerning the scalar potential $V(\phi)$ the usual assumption in
the literature is to assume an exponential potential of the form
\begin{equation}
\label{exppot}
V=V_0\exp(-\kappa\lambda\phi),
\end{equation}
since exponential potentials are known to be significant in
various cosmological models
\cite{Schmidt:1990gb,Muller:1989rp,Copeland:1997et,Ferreira:1997au,
Gong:2006sp,Chen:2008pz,Chen:2008ft} (equivalently, but more
generally, we could consider potentials satisfying
$\lambda=-\frac{1}{\kappa
V(\phi)}\frac{\partial V(\phi)}{\partial\phi}\approx const$, which is valid
for arbitrary but nearly flat potentials \cite{Scherrer:2007pu,Scherrer:2008be,Setare:2008sf}).
Moreover, note that the exponential potential was used as an example in
the initial work on teleparallel dark energy
\cite{Geng:2011aj}.

In summary, using the auxiliary variables (\ref{auxiliary}) and
considering the exponential potential (\ref{exppot}),
the equations of motion (\ref{FR1}), (\ref{FR2}) and (\ref{fieldevol2}) in
the case of dust matter,
can be transformed into the following autonomous system:

\begin{align}
 &x'=\frac{3 x^3}{2  z^2 \text{sgn}(\xi )+2}+\frac{2 \sqrt{6} z
   \sqrt{|\xi |} x^2\text{sgn}(\xi )}{  z^2\text{sgn}(\xi )+2}-\frac{3
x}{2} 
+\frac{1}{2} y^2
   \left[\sqrt{6} \lambda -\frac{3 x}{\text{sgn}(\xi )
   z^2+1}\right]\nonumber\\
&\ \ \ \ \ \ \
-\sqrt{6} z \sqrt{|\xi|} \text{sgn}(\xi)\nonumber\\
&y'=\frac{3 y x^2}{2
   \text{sgn}(\xi ) z^2+2}+\frac{3}{2} y
\left[1-\frac{y^2}{ z^2\text{sgn}(\xi )
  +1}\right]
-\frac{\sqrt{\frac{3}{2}} yx
\left\{\left[\lambda  z-4   \sqrt{|\xi |}\right] z\text{sgn}(\xi
)+\lambda\right\} }{1+
   z^2\text{sgn}(\xi )}\nonumber\\
&z'=\sqrt{6} \sqrt{|\xi |} x,
\label{autonomous}
\end{align}
where we have used that
for every
quantity $F$ we acquire $\dot{F}=HF'$.
Since $\rho_m$ is
nonnegative, from (\ref{Fr1norm}) we obtain that  $x^2+y^2-z^2
\text{sgn}(\xi )\leq 1$.
Thus, we deduce that for $\xi<0$ the system \eqref{autonomous}
defines a flow on the compact phase space
$\Psi=\left\{ x^2+y^2-z^2 \text{sgn}(\xi )\leq 1, y\geq 0\right\},$ for
$\xi=0$ the phase space is compact and it is reduced to the circle
$\Psi=\left\{ x^2+y^2\leq 1, y\geq 0\right\},$  while for
$\xi>0$ the phase space $\Psi$ is unbounded.

Before proceeding we make two comments on the degrees of freedom and the
choice of auxiliary variables. Firstly, as we have already mentioned,
in the minimal coupling case, that is when $\xi=0$, the model at hand
coincides with standard quintessence, which phase space analysis is well
known using two degrees of freedom, namely the variables $x$ and $y$
defined above \cite{Copeland:1997et}. On the other hand, when $\xi\neq0$ we have
three degrees of freedom and all $x$, $y$, $z$ are necessary. Therefore, in
order to perform the analysis in a unified way, we use the three variables
defined above, having in mind that for $\xi=0$ the variable $z$ becomes
zero and thus irrelevant. Secondly,
apart from the standard choices
of the variables $x$ and $y$, one must be careful in suitably choosing the
variable $z$ in order not to lose dynamical information. For example,
although the choice for the variable $z=\frac{\kappa\rho_m}{\sqrt{3}H}$ is
another reasonable choice, however it still loses the critical points that
lie at ``infinity''. Therefore, in order to completely cover the
phase-space behavior in the following we will additionally use the
Poincar\'e central projection method to investigate the dynamics at
``infinity''. The negligence of this point was the reason of the incomplete
phase space analysis of teleparallel dark energy performed in
\cite{Wei:2011yr}.

\subsection{Finite Phase-space analysis}
\label{phasespacefinite}

Let us now proceed to the phase-space analysis. The real and physically
meaningful (that is corresponding to an expanding universe, i.e possessing
$H>0$) critical points $(x_c, y_c, z_c)$ of the autonomous system
(\ref{autonomous}), obtained by setting the left hand sides of the
equations to zero, are presented in Table \ref{crit}. In the same table we
provide their existence conditions. The $3\times3$ matrix ${\bf
{Q}}$ of the linearized perturbation equations of the system
(\ref{autonomous}) is shown in   Appendix \ref{appaa}. For each critical
point of
Table \ref{crit} we examine the sign of the real part of the eigenvalues of
${\bf {Q}}$ in order to determine the type and stability of the point. The
details of the analysis and the various eigenvalues are presented in
Appendix  \ref{appaa}, and in Table \ref{crit1} we summarize the results.
 \footnote{The critical point $D$ that exists only for $\xi=0$, is very
close to be a stationary point in the general case ($\xi\neq 0$) too,
but with an arbitrarily small value of the slope of the potential
$\lambda$. It becomes an exact critical point only for $\lambda=0$, namely
the point $J$. Furthermore, for arbitrary large values of $\lambda$ the
singular point $E$ (that exists for $\xi= 0$) mimics the behavior of the
point $A$ (that exists for $\xi\neq 0$).} In addition,
for each critical point we calculate the values of $\Omega_{DE}$,
$w_{DE}$, $w_{tot}$ and $q$  given by (\ref{Omegas}), (\ref{wdephase}),
(\ref{wtot}) and (\ref{decc}) respectively.

 \begin{table*}[!]
\begin{center}
\begin{tabular}{|c|c|c|c|c|}
\hline
&&&&   \\
 Cr. P.& $x_c$ & $y_c$ & $z_c$ & Exists for \\
\hline \hline
$A$& 0 & 0 & 0 &  all $\xi$,$\lambda$
\\
\hline
 $B$& 1& 0 & 0 &  $\xi=0$, all $\lambda$ \\
\hline
 $C$& -1& 0 & 0 &  $\xi=0$, all $\lambda$ \\
\hline
$D$& $\frac{\lambda}{\sqrt{6}}$ & $\sqrt{1-\frac{\lambda^2}{6}}$ &
0
&  $\xi=0$,   $ \lambda^2\leq6$ \\
\hline
 $E$& $\sqrt{\frac{3}{2}}\frac{1}{\lambda}$ &
$\sqrt{\frac{3}{2}}\frac{1}{\lambda}$ &
0 &  $\xi=0$,    $\lambda^2\geq3$ \\
\hline
 $F$& 0 & $\sqrt{\frac{2\xi -2\sqrt{\xi  \left(\xi -\lambda
^2\right)}}{\lambda ^2}}$ &
$\frac{\left[\xi -\sqrt{\xi  \left(\xi -\lambda ^2\right)}\right]
\sqrt{|\xi |}}{\lambda  \xi }$& $0<\lambda^2\leq\xi$ \\
\hline
 $G$&0 & $\sqrt{\frac{2\xi +2\sqrt{\xi  \left(\xi -\lambda
^2\right)}}{\lambda ^2}}$ &
$\frac{\left[\xi +\sqrt{\xi  \left(\xi -\lambda ^2\right)}\right]
\sqrt{|\xi |}}{\lambda  \xi }$ & $0<\lambda^2\leq\xi$ or
$\xi<0$ \\
\hline
$J$&0 & 1 & 0& $\xi\neq0$ and $\lambda=0$ \\
\hline
\end{tabular}
\end{center}
\caption[crit]{\label{crit} The real and physically meaningful
critical points of the autonomous system (\ref{autonomous}). Existence
conditions.}
\end{table*}

 \begin{table*}[!]
\begin{center}
\begin{tabular}{|c|c|c|c|c|c|}
\hline
&&&&&   \\
 Cr. P.& 
Stability & $\Omega_{DE}$ &  $w_{DE}$ & $w_{tot}$ & $q$\\
\hline \hline
$A$&  saddle point &   0 & arbitrary & 0 &$\frac{1}{2}$ \\
\hline
 $B$&   unstable for
$\lambda<\sqrt{6}$ &
   &     &   & \\
  &    saddle point otherwise &
1  &   1 & 1 & 2\\
\hline
 $C$&   unstable for
$\lambda>-\sqrt{6}$ &
   &     &   & \\
  &   saddle point otherwise &
1  &   1 & 1 & 2\\
\hline
$D$& stable node for $\lambda^2<3$ &   1 &
$-1+\frac{\lambda^2}{3}$ & $-1+\frac{\lambda^2}{3}$ &
$-1+\frac{\lambda^2}{2}$\\
& saddle point for $3<\lambda^2<6$&    &  & &\\
\hline
 $E$& stable
node for
 $3<\lambda^2<\frac{24}{7}$ & $\frac{3}{\lambda^2}$ & 0 & 0 &
$\frac{1}{2}$\\
& stable spiral for
 $\lambda^2>\frac{24}{7}$&    &  & &\\
\hline
 $F$&
stable node for  $\lambda^2<\xi$ &  1 & $-1$ & $-1$ & $-1$\\
\hline
 $G$& saddle point &  1 & $-1$ & $-1$ & $-1$\\
\hline
$J$& stable spiral for
$\frac{3}{8} <\xi$ & 1 & $-1$ & $-1$ & $-1$\\
&    stable node for
$0<\xi<\frac{3}{8}$
&    &  & &\\
&  saddle point for
 $\xi<0$
&    &  & &\\
\hline
\end{tabular}
\end{center}
\caption[crit]{\label{crit1} The real and physically meaningful
critical points of the autonomous system (\ref{autonomous}). Stability
conditions, and the
corresponding values of the dark-energy
density parameter $\Omega_{DE}$, of the  dark-energy equation-of-state
parameter $w_{DE}$, of the total equation-of-state parameter
  $w_{tot}$ and of the deceleration parameter $q$. }
\end{table*}

\subsection{Phase-space analysis at infinity}
\label{phasespaceinfinite}

\begin{table*}[!]
\begin{center}
\begin{tabular}{|c|c|c|c|c|c|c|c|c|}
\hline
&&&&&&&& \\
 Cr. P. & $x_{rc}$ & $y_{rc}$ & $z_{rc}$  &
Stability & $\Omega_{DE}$ &  $w_{DE}$ & $w_{tot}$ & $q$\\
\hline \hline
$K_\pm$& $\mp\frac{\sqrt{2}}{2}$ & 0 &
$\pm\frac{\sqrt{2}}{2}$
& unstable for &  &  & & \\ &  &  & 
& $0<\xi <\frac{3}{8}$    &arbitrary & arbitrary  &  $1-4
\sqrt{\frac{2\xi}{3}} $ &$2-2 \sqrt{6\xi}  $  \\   &  &  &
&   saddle point &  &  & & \\ &  &  & 
& for $\frac{3}{8}<\xi$ &  & &&
\\
\hline
$L_{\pm}$& $\pm\frac{\sqrt{2}}{2}$ & 0 &
$\pm\frac{\sqrt{2}}{2}$
& saddle point &  &  & & \\ &  &  & 
& for all $\xi>0$   & arbitrary & arbitrary & $1+4
\sqrt{\frac{2\xi}{3}}
 $&
$2+2 \sqrt{6\xi} $ \\
\hline
\end{tabular}
\end{center}
\caption[crit]{\label{crit2} The real and physical critical points of the
autonomous system (\ref{autonomous}) at infinity, which exists for $\xi>0$,
all $\lambda$, and the
corresponding values of the dark-energy
density parameter $\Omega_{DE}$, of the  dark-energy equation-of-state
parameter
$w_{DE}$, of the total equation-of-state parameter
  $w_{tot}$ and of the deceleration parameter $q$.   }
\end{table*}

Due to the fact that the dynamical system
\eqref{autonomous} is non-compact for the choice $\xi>0$, there could be
features in the asymptotic regime which are non-trivial for the
global dynamics. Thus, in order to complete the analysis of the
phase space we must extend our study using the Poincar\'e
central projection method \cite{PoincareProj}.

Let us consider the Poincar\'e variables
\begin{equation}\label{Transf0}
x_r=\rho \cos\theta \sin \psi ,\ z_r=\rho
\sin \theta \sin \psi ,\, y_r= \rho \cos \psi,\,
\end{equation}
where $\rho=\frac{r}{\sqrt{1+r^2}},$ $r=\sqrt{x^2+y^2+z^2},$
$\theta\in[0,2\pi],$
and $-\frac{\pi}{2}\leq \psi \leq \frac{\pi}{2}$ (we
restrict the angle $\psi$ to this interval since the physical
region is given by $y>0$) \cite{PoincareProj,Leon:2008de,Leon:2010ai,Leon2011}. Thus, the points at ``infinity''
($r\rightarrow+\infty$) are those having $\rho\rightarrow 1$. Furthermore,
the physical phase space is given by
$$\left\{\left(x_r,y_r,z_r\right)\in [-1,1]\times [0,1]\times[-1,1]| z_r^2\leq x_r^2+y_r^2\leq
\frac{1}{2}\right\}.$$

Inverting relations (\ref{Transf0}) and substituting into
(\ref{Omegas}),(\ref{wdephase}), we obtain the dark energy density and
equation-of-state parameters as a function of the  Poincar\'e variables,
namely:
\begin{eqnarray}
&\Omega_{DE}=\frac{x_r^2+y_r^2-z_r^2}{1-x_r^2-y_r^2-z_r^2}
 \label{Omegas22}\\
&w_{DE}=\frac{\left(3 x_r^2+4 \sqrt{6} \sqrt{\xi } z_r x_r-3 y_r^2\right) \left(1-x_r^2-y_r^2-z_r^2\right)}{3 \left(1-x_r^2-y_r^2\right)
   \left(x_r^2+y_r^2-z_r^2\right)},
\label{wdephase22}
\end{eqnarray}
and similarly substituting into
(\ref{wtot}), (\ref{decc})  we obtain the corresponding expressions for
the total equation-of-state and deceleration parameters:
\begin{eqnarray}
w_{tot}
=\frac{3 x_r^2+4 \sqrt{6} \sqrt{\xi } z_r x_r-3 y_r^2}{3 \left(1-x_r^2-y_r^2\right)},
\label{wtotphase22}
\end{eqnarray}
\begin{eqnarray}
q=\frac { 1 } { 2 }
+\frac{3}{2}\left\{\frac{3 x_r^2+4 \sqrt{6} \sqrt{\xi } z_r x_r-3 y_r^2}{3 \left(1-x_r^2-y_r^2\right)}\right\}.
\label{wqphase22}
\end{eqnarray}

Applying the procedure prescribed in Appendix \ref{Poincare} we conclude
that there are four physical critical points at infinity, namely the
points $K_\pm$, satisfying $\cot\theta=-1$, and the points $L_\pm$
satisfying $\cot\theta=1$. These critical points,
along with their stability conditions are presented
in Table \ref{crit2}. In the same Table we include the corresponding
values of the observables  $\Omega_{DE}$, $w_{DE}$, $w_{tot}$ and $q$,
calculated using
(\ref{Omegas22}),(\ref{wdephase22}),(\ref{wtotphase22}),(\ref{wqphase22}).

As we show in Appendix \ref{Poincare}, the above critical points
correspond to the limit
\begin{equation}\label{inftybehavior} |\phi|\rightarrow \infty,\
\left|\frac{\dot\phi}{H}\right| \rightarrow \infty,\end{equation}
satisfying the rate
$\left(\ln\phi\right)'\equiv
\sqrt{6}\frac{x}{z}=\sqrt{6}\cot\theta$.

\section{Cosmological Implications} \label{implications}

Having performed the complete phase-space analysis of teleparallel dark
energy, we can now discuss the corresponding cosmological behavior. A
first remark is that in the minimal case (that is $\xi=0$) we do verify
that the scenario at hand coincides completely with standard quintessence.
Therefore, we will make a brief review on the subject and then focus
on the nonminimal case.

The points $B$ to $E$ exists only for the minimal case, that is only for
$\xi=0$. Points $B$ and $C$ are not stable,
corresponding to a non-accelerating, dark-energy dominated universe, with a
stiff dark-energy equation-of-state parameter equal to 1. Both of them
exist in standard quintessence \cite{Copeland:1997et}.

Point $D$ is a saddle one for $3<\lambda^2<6$, however for $0<\lambda^2<3$
it is a stable node, and thus it can attract the universe at late times.
It corresponds to a dark-energy dominated universe, with a dark-energy
equation-of-state parameter lying in the quintessence regime, which can be
accelerating or not according to the $\lambda$-value. This point
exists in standard quintessence \cite{Copeland:1997et}. It is  the most
important one in that scenario, since it is both stable and possesses a
$w_{DE}$ compatible with observations.

Point $E$ is a stable one for $3<\lambda^2$. It has the advantage that the
dark-energy density parameter lies in the interval $0<\Omega_{DE}<1$, that
is it can alleviate the coincidence problem, since dark matter and dark
energy density parameters can be of the same order (in order to treat the
coincidence problem one must explain why the present dark energy and matter
density parameters are of the same order of magnitude although they follow
different evolution behaviors). However, it has the disadvantage that
$w_{DE}$ is 0 and the expansion is not accelerating, which are not favored
by observations. This point exists in standard quintessence \cite{Copeland:1997et}.

Let us now analyze the case $\xi\neq0$. In this case we obtain the
critical points $A$, $F$, $G$ and $J$. The point $A$ is saddle point, and
thus it cannot be late-time solution of the universe. It corresponds to a
non-accelerating, dark-matter dominated universe, with arbitrary
dark-energy equation-of-state parameter. Note that this trivial point
exists in the standard quintessence model too \cite{Copeland:1997et}, since it is
independent of $\xi$.

The present scenario of teleparallel dark energy, possesses two
additional, non-trivial critical points that do not exist in standard
quintessence. Thus, they account for the new information of this richer
scenario, and as expected they depend on the nonminimal coupling $\xi$. In
particular, point $F$ is stable if $\lambda^2<\xi$, and thus it can attract
the universe at late times. It corresponds to an accelerating
universe with complete dark energy domination, with $w_{DE}=-1$, that is
dark energy behaves like a cosmological constant. We stress that this
$w_{DE}$ value is independent of $\lambda$ and $\xi$, which is an important
and novel result. Thus, while point $D$ (the important point of standard
quintessence) needs to have  a very flat, that is quite tuned, potential in
order to possess a $w_{DE}$ near the observed value $-1$, point $F$
exhibits this behavior for every $\lambda$-value, provided that
$\lambda^2<\xi$. This feature is a significant advantage of the scenario at
hand, amplifying its generality, and offers a mechanism for the
stabilization of $w_{DE}$ close to the cosmological-constant value.
Similarly, we have the   point $G$, which has the same cosmological
properties with $F$, however it is a saddle point and thus it cannot be a
late-time solution of the universe, but the universe can spend a
large period of time near this solution.

Finally, when $\xi\neq0$ and for the limiting case $\lambda=0$, that is for
a constant potential, the present scenario exhibits the critical point $J$,
which is stable for $\xi>0$. It corresponds to a dark-energy dominated, de
Sitter universe, in which dark energy behaves like a cosmological
constant.

Let us now now analyze the critical points at infinity. The points $K_\pm$
are saddle for $\xi >\frac{3}{8}$. The fact that they possess
arbitrary $\Omega_{DE}$ and $w_{DE}$ is a significant advantage, since such
solutions can alleviate the coincidence problem. Note however that the
corresponding $w_{tot}$ and deceleration parameter are not arbitrary,
offering a good quantitative description of the cosmological behavior
(actually this is the reason we introduced these observables). In
particular, we conclude that for $\xi >\frac{1}{6}$
we obtain an accelerated expansion, while for $\xi >\frac{3}{8}$ the
expansion is super-accelerating ($q<-1$ that is $\dot{H}>0$ and $w_{tot}<-1$) that is the
universe presents a phantom behavior. We mention that since these points
are saddle they cannot attract the universe at
late times, however the universe can spend a large period of time near
these solutions, before approaching the saddle point $A$    or the global
attractor $F$ (for
potentials with slope $\lambda^2\leq\xi$). In that case the above physical
features present a transient character, which is quite significant from
the observational point of view.

Points $L_\pm$ are saddle for all $\xi>0$. They have arbitrary
$\Omega_{DE}$ and $w_{DE}$ however they correspond to a non-accelerating
expansion, and thus they could be important only as transient states of the
universe.

\begin{figure}[ht]
\begin{center}
\includegraphics[height=8.4cm,width=9.4cm]{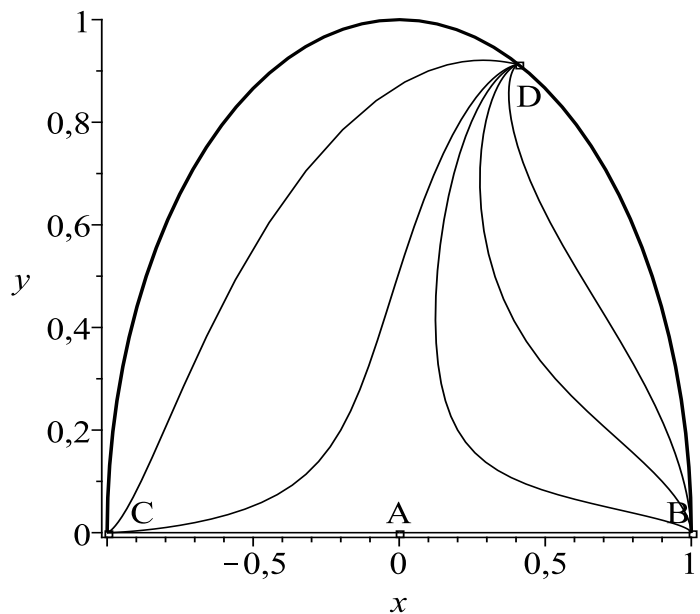}
\caption{{\it{The projection of the phase-space evolution on the $x-y$
plane,
for the teleparallel dark energy scenario with $\lambda=1$ and $\xi=0$.
The trajectories are attracted by the quintessence-like  stable point $D$,
which exists only in the minimal case ($\xi=0$). }} }
\label{fig3}
\end{center}
\end{figure}

In order to present the aforementioned behavior more
transparently, we first evolve the autonomous system (\ref{autonomous})
numerically for the choice  $\lambda=1$, and $\xi=0$, that is in the
minimal scenario, and in
Fig. \ref{fig3} we depict the corresponding phase-space behavior,
projected in the $x-y$ plane. As we can wee, in this case the
quintessence-like critical point $D$ is the late-time solution of the
universe.
\begin{figure}[!]
\begin{center}
\includegraphics[height=8.4cm,width=9.4cm]{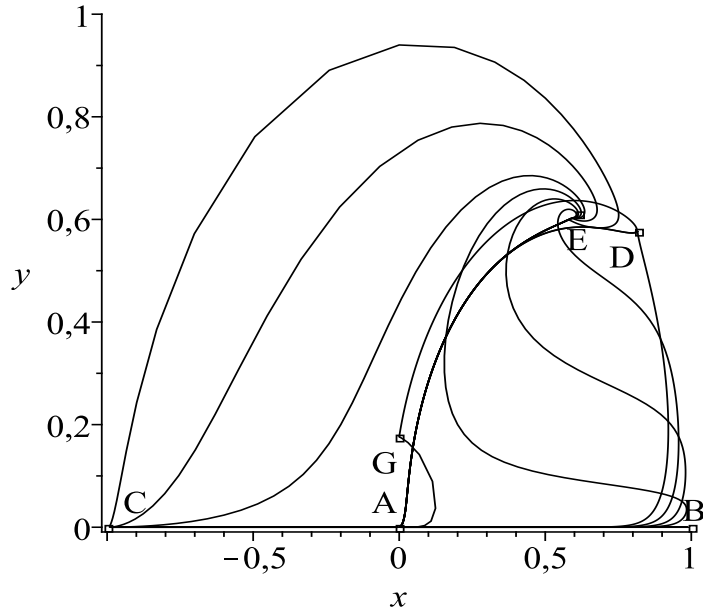}
\caption{{\it{The projection of the phase-space evolution on the
$x-y$ plane, for the teleparallel dark energy scenario with
$\lambda=2$ and $\xi=-10^{-3}$. Observe that an open set of
trajectories remains very close to the quasi-stationary solution
$E$, before approaching an invariant arc above the saddle point
$A$. The critical point $G$ is a saddle one.}} }
 \label{fig1a}
\end{center}
\end{figure}
Note that this point exists only for the minimal case
($\xi=0$), where our model coincides with standard quintessence (see Fig.
2 of \cite{Copeland:1997et}).
\begin{figure}[ht]
\begin{center}
\includegraphics[height=10.3cm,width=11.3cm]{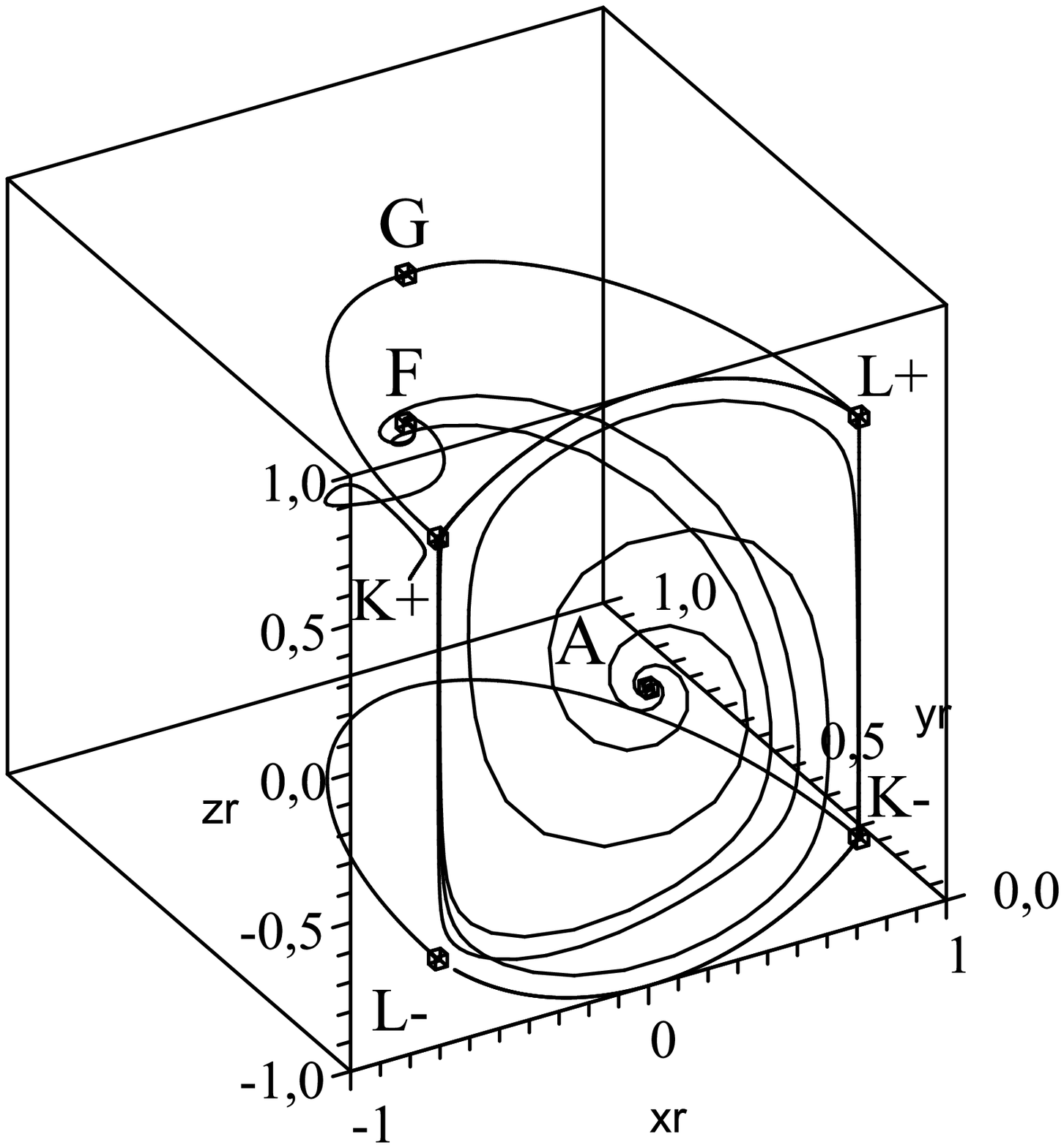}
\caption{{\it{Poncar\'e (global) phase space for the teleparallel
dark energy scenario with $\lambda=0.5$ and $\xi=1$. The points at
infinity $K_\pm, L_\pm$ are saddles. The trajectories on both
finite and infinite regions are finally attracted by the
cosmological-constant-like stable point $F$ (global attractor), having
passed through matter-dominated solutions or solutions with
comparable dark matter and dark energy sectors.}}
} \label{fig5}
\end{center}
\end{figure}

As we mentioned above, the late-time attractors $D$ and $E$ exists only for
the case $\xi=0$. To illustrate how sensible is the scenario at hand in
slight changes of $\xi$, in Fig. \ref{fig1a} we depict the projection of
the phase-space evolution on the $x-y$ plane, for $\lambda=2$ and
$\xi=-10^{-3}$. In this case the point $E$ becomes ``quasi-stationary'',
since typical trajectories remain very close to it for a large but finite
time interval, while $D$ becomes also quasi-stationary, but it is unstable
to perturbations in the $z$-axis. \footnote{Although the singular points
$B,C, D$ and $E$ do not exist as critical points for the choices
$\xi\neq 0$, in Fig.  \ref{fig1a} we mark their locations explicitly, for
comparison with the case $\xi=0$ of Fig. \ref{fig3}.}

Finally, to illustrate the dynamics for positive $\xi$, which is expressed
through points at ``infinity'',  in Fig. \ref{fig5} we present the
three-dimensional Poncar\'e (global) phase space for $\lambda=0.5$ and
$\xi=1$.

 According to Table \ref{crit2} the critical points   $K_\pm$ are saddle
ones and exhibit phantom behavior, while  $L_\pm$ are saddle with a stiff
dark energy sector. Thus, far from their basin of attraction (but inside
the invariant set $y_r=0$) the orbits departs from them to approach the
matter-dominated solution $A$, while for $y_r>0$ the orbits are attracted
by the cosmological-constant-like solution $F$. Thus, the epoch sequence
$K_-\rightarrow L_+\rightarrow K_+\rightarrow L_-\rightarrow
A\rightarrow F$ represents the transition from  a universe
with a phantom-like dark energy, to a matter-dominated universe with
non-phantom dark energy, and then to a dark-energy-dominated,
cosmological-constant-like solution. This sequence has a great cosmological
significance, since it can describe the epoch sequence inflation,
dark-matter domination, dark-energy domination, in agreement with
observations. Additionally, the epoch sequence $L_-\rightarrow
K_+\rightarrow F$ represents the transition from a universe with
comparable dark matter and dark energy sectors, to a
cosmological-constant-like solution.
\begin{figure}[ht]
\begin{center}
\includegraphics[height=10.3cm,width=11.3cm]{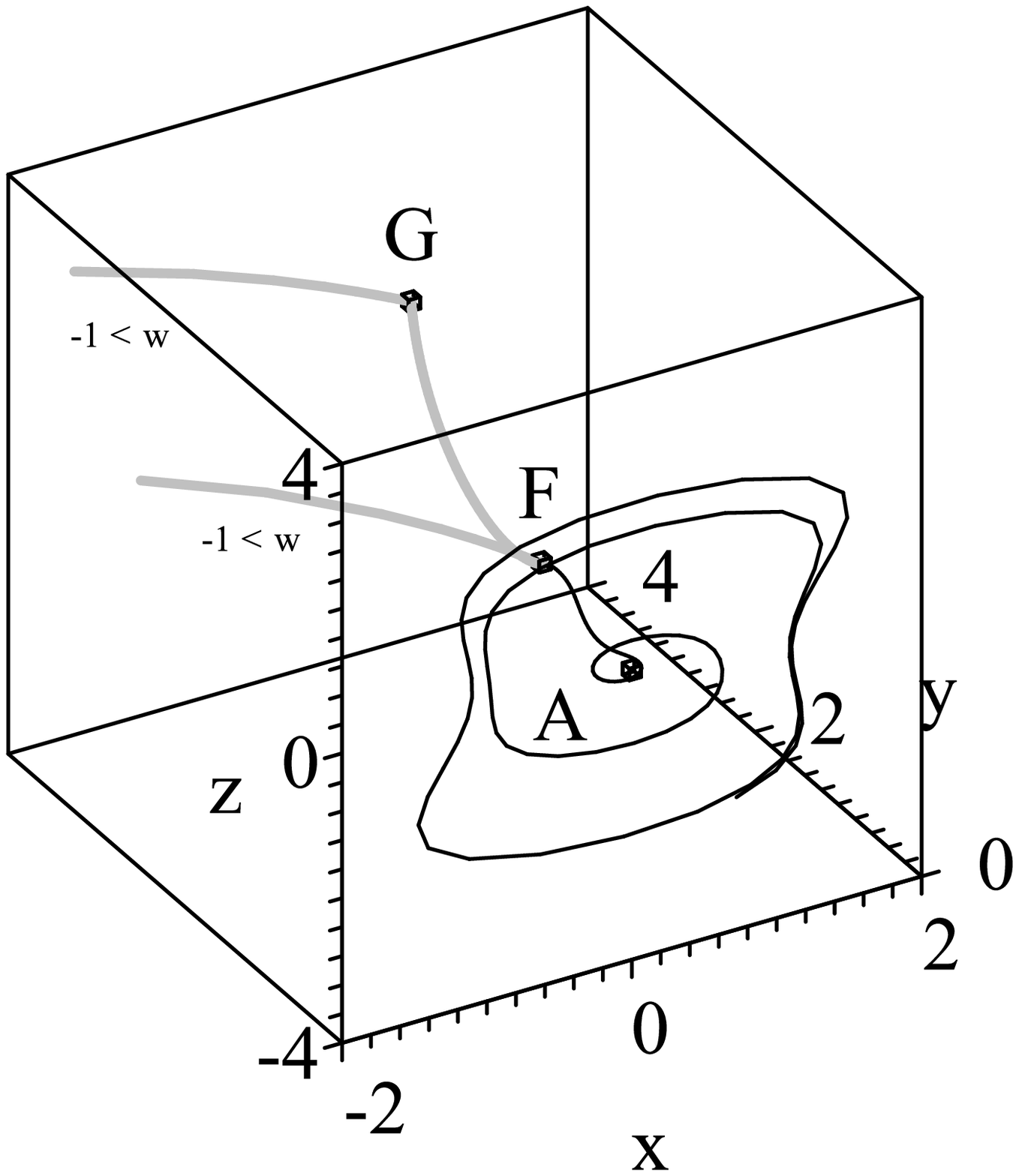}
\caption{{\it{The phase-space evolution for the teleparallel dark
energy scenario with $\lambda=0.7$ and $\xi=1$. The trajectories
are attracted by the cosmological-constant-like stable point $F$,
however the evolution towards it possesses a $w_{DE}$ being
quintessence-like, phantom-like, or experiencing the
phantom-divide crossing, depending on the specific initial
conditions. The orbits with a thick gray curve on the top left are those
with $w_{DE}>-1$ initially, while the thin black curve corresponds to
$w_{DE}<-1$ initially.}} } \label{fig2}
\end{center}
\end{figure}

Before closing this section, let us make a comment on another crucial
difference of teleparallel dark energy, comparing with standard
quintessence, that is of the $\xi\neq0$ case comparing to the $\xi=0$ one.
In particular, when $\xi=0$, in which teleparallel dark energy coincides
with standard quintessence, $w_{DE}$ is always larger that $-1$, not only
at the critical points, but also throughout the cosmological evolution as
well. However, for $\xi\neq0$, during the cosmological evolution $w_{DE}$
can be either above or below $-1$, and only at the stable critical point
it becomes equal to $-1$. Such a cosmological behavior is much
richer, and very interesting, both from the theoretical as well as from
the observational point of view, since it can explain the dynamical
behavior of $w_{DE}$ either above or below the phantom divide, and
moreover its stabilization to the cosmological-constant value.

In order to present the novel features of the scenario at hand, in Fig.
\ref{fig2} we depict the three-dimensional phase-space behavior for the
choice $\lambda=0.7$ and $\xi=1$.
In this case the universe at late times is attracted  by the
cosmological-constant-like stable point $F$. However, during the evolution
towards this point the dark-energy equation-of-state parameter $w_{DE}$
presents a very interesting behavior, and in particular, depending on the
specific initial conditions, it can be quintessence-like, phantom-like, or
experience the phantom-divide crossing. Such a behavior is much richer
than standard quintessence, and reveals the capabilities of teleparallel
dark energy scenario. Finally, one should also investigate in detail
whether the scenario at hand is stable  at the perturbation level,
especially in the region $w_{DE}<-1$. Such an analysis can be heavily based
on \cite{Chen:2010va} and one can find that  $f(T)$ cosmology and
teleparallel dark energy are stable even in the region $w_{DE}<-1$. This
project is in preparation \cite{inprep}.

\section{Conclusions}
\label{Conclusions}

In the present work we investigated the dynamical behavior of the recently
proposed scenario of teleparallel dark energy
\cite{Geng:2011aj,Geng:2011ka},
which is based on the teleparallel equivalent of General Relativity
(TEGR), that is on its torsion instead of curvature formulation
\cite{Unzicker:2005in,Hayashi:1979qx}. In this model one adds a
canonical scalar
field, in which the dark energy sector is attributed, allowing also for a
nonminimal coupling between the field and the torsion scalar. Thus,
although the minimal case is completely equivalent with
standard quintessence, the nonminimal scenario has a richer structure,
exhibiting quintessence-like or phantom-like behavior, or experiencing the
phantom-divide crossing \cite{Geng:2011aj,Geng:2011ka}.

Performing a detailed phase-space analysis of teleparallel dark energy, we
saw that the standard quintessence-like, late-time
solution \cite{Copeland:1997et} exists only for $\xi=0$, and we showed that this
solution becomes quasi-stationary if $\xi$ is perturbed even slightly.
The same results hold for the other standard quintessence solution,
namely the stiff dark-energy late-time attractor, in which dark energy and
dark matter can be of the same order, thus alleviating the coincidence
problem.

Apart from the above late time solutions that exist also in the
standard quintessence scenario, teleparallel dark energy has an additional
and physically very interesting late-time behavior. In particular, it
possesses a late-time attractor in which dark energy behaves like a
cosmological constant, independently of the specific values of the model
parameters (provided that the nonminimal coupling $\xi$ is larger than
the square of the potential exponent). This feature is a significant
advantage of the scenario at hand, amplifying its generality, since it
provides a natural way for the stabilization of the dark energy
equation-of-state parameter to the cosmological constant value, without
the need for parameter-tuning.

Additionally, we showed that teleparallel dark energy admits critical
points at ``infinity'', which are saddle for $\xi>\frac{3}{8}$, and thus although they cannot be
the late-time solutions, the universe can spend a large period of time near
them. Thus, one can obtain the transition from a universe
with a phantom-like dark energy, to a matter-dominated universe with
non-phantom dark energy, and then to a dark-energy-dominated,
cosmological-constant-like solution. This sequence has a great cosmological
significance, since it can describe the epoch sequence inflation,
dark-matter domination, dark-energy domination, in agreement with
observations

Finally, perhaps the most interesting feature of teleparallel dark energy
is that in the nonminimal case the dark energy equation-of-state parameter
can be either above or below $-1$, and only at the stable critical points
it becomes equal to $-1$. Such a cosmological behavior is much
richer comparing to standard quintessence, and very interesting, both from
the theoretical as well as from the observational point of view, since it
can explain the dynamical behavior of $w_{DE}$ either above or below the
phantom divide, and moreover its stabilization to the cosmological-constant
value. All these features make the teleparallel dark energy a good
candidate for the description of  dark energy.

\begin{acknowledgments}
The authors would like to thank Yungui Gong for crucial discussions and
comments. This work was partially supported by the NNSF key project of
China under grant No. 10935013, the National Basic Science Program (Project
973) of China under grant No. 2010CB833004, CQ CSTC under grant No.
2009BA4050 and CQ CMEC under grant No. KJTD201016. ENS
wishes to thank C.M.P. of Chongqing University of Posts and
Telecommunications, for the hospitality during the preparation of the
manuscript. GL would like to thank the MES of Cuba for partial
financial support of this investigation, and his research was also
supported by the National Basic Science Program (PNCB) of Cuba and
Territorial CITMA Project (No. 1115).
\end{acknowledgments}

\begin{appendix}

\section{Stability of the finite critical points}
\label{appaa}

\begin{table*}[htp]
\begin{center}
\begin{tabular}{|c|c|c|c|c|}
\hline
 Cr. Point & Exists for & $\nu_1$ & $\nu_2$ & $\nu_3$
\\
\hline \hline
 A & all $\xi$,$\lambda$  & $\frac{3}{2}$ & $ \frac{1}{4}
\left(-3-\sqrt{9-96 \xi }\right)$ & $\frac{1}{4} \left(-3+\sqrt{9-96 \xi
}\right)$ \\
\hline
 B & $\xi=0$, all $\lambda$ & $3$ & $3-\sqrt{\frac{3}{2}}\lambda$ & 0 \\
\hline
C & $\xi=0$, all $\lambda$ & $3$ & $3+\sqrt{\frac{3}{2}}\lambda$ & 0 \\
\hline
D& $\xi=0$,  $\lambda^2\leq6$ & $\lambda^2-3$ & $\frac{1}{2} \left(\lambda
^2-6\right)$ &
0   \\
\hline
 E   & $\xi=0$,  $\lambda^2\geq3$ & $\alpha^+(\lambda)$ &
$\alpha^-(\lambda)$  &0 \\
\hline
F & $\lambda^2\leq\xi$  & $-3$ &
$\beta^+(\lambda,\xi)$ &
$\beta^-(\lambda,\xi)$ \\
\hline
G  & $ \lambda^2\leq\xi$
or $\xi<0$ & $-3$ & $\gamma^+(\lambda,\xi)$ &
$\gamma^-(\lambda,\xi)$\\
\hline
J & $\xi\neq0$
and $\lambda=0$ & $-3$ &
$\frac{-3+\sqrt{9-24 \xi }}{2}$ &
$ \frac{-3-\sqrt{9-24\xi}}{2}$\\
\hline
\end{tabular}
\end{center}
\caption[stability]{\label{eigen} The eigenvalues of the matrix
${\bf {Q}}$ of the perturbation equations of the autonomous system
(\ref{autonomous}). Points $B$-$E$ exist only for $\xi=0$, and for
these points the variable $z$ is zero and thus irrelevant. Although for
 these points the eigenvalue associated to the z-direction is zero,
the stability conditions are obtained by analyzing the eigenvalues of the
non-trivial $2\times2$ submatrix of  ${\bf {Q}}$. We introduce the notations $\alpha^\pm(\lambda)=\frac{3}{4} \left(-1\pm\frac{\sqrt{24
\lambda ^2-7 \lambda ^4}}{\lambda ^2}\right),$ $\beta^\pm(\lambda,\xi)=\frac{-3\pm\sqrt{9-24\sqrt{\xi^2-\lambda^2\xi}}}{2}$ and $\gamma^\pm(\lambda,\xi)=\frac{-3\pm\sqrt{9+24\sqrt{\xi^2-\lambda^2\xi}}}{2}.$}
\end{table*}
For the critical points $(x_c,y_c,z_c)$ of the autonomous system
(\ref{autonomous}), the coefficients of the perturbation equations form a
$3\times3$ matrix ${\bf {Q}}$, which reads:

\begin{eqnarray}
{\bf
{Q}}_{11}&=&\frac{9 x^2-3 y^2}{2+2 z^2 \text{sgn}(\xi ) }+\frac{4 \sqrt{6}
x z    \sqrt{|\xi |}\text{sgn}(\xi )}{2+ z^2\text{sgn}(\xi )}-\frac{3}{2}
\nonumber\\
{\bf
{Q}}_{12}&=&y \left[\sqrt{6} \lambda -\frac{3 x}{1+z^2\text{sgn}(\xi )
}\right]
\nonumber\\
{\bf
{Q}}_{13}&=&-\frac{2 \sqrt{6}    \left[z^2 \text{sgn}(\xi )-2 \right]
\sqrt{|\xi |} x^2\text{sgn}(\xi )}{\left[2+ z^2\text{sgn}(\xi
)\right]^2}+\frac{3
   \left(y^2-x^2\right) z x\,\text{sgn}(\xi ) }{\left[1+z^2\text{sgn}(\xi )
\right]^2}-
\sqrt{6} \sqrt{|\xi |}\text{sgn}(\xi )
\nonumber\\
{\bf
{Q}}_{21}&=&\frac{3 x y}{1+z^2\text{sgn}(\xi )}-\frac{\sqrt{\frac{3}{2}} y
\left[\lambda   z^2\text{sgn}(\xi )-4 z    \sqrt{|\xi |} \text{sgn}(\xi
)+\lambda  \right]}{1+ z^2\text{sgn}(\xi )}
\nonumber\\
{\bf
{Q}}_{22}&=&
\frac{3+3 x^2-\sqrt{6} \lambda  x-9 y^2}{2+2 z^2\text{sgn}(\xi ) }
+\frac{
\left\{3 z+\sqrt{6} x \left[4 \sqrt{|\xi |}-
\lambda z\right]\right\}z\,\text{sgn}(\xi ) }{2+2
z^2\text{sgn}(\xi ) }
\nonumber\\
{\bf
{Q}}_{23}&=&\frac{y \,\text{sgn}(\xi )  \left\{3 z\left(y^2-x^2\right)
+2 x \sqrt{6}  \sqrt{|\xi|}\left[1-z^2\text{sgn}(\xi )
\right] \right\}}{\left[1+z^2
   \text{sgn}(\xi )\right]^2}
\nonumber\end{eqnarray}
\begin{eqnarray}
{\bf
{Q}}_{31}&=&\sqrt{6} \sqrt{|\xi |}\nonumber\\
{\bf
{Q}}_{32}&=&0
\nonumber\\
{\bf
{Q}}_{33}&=&0.\nonumber
\end{eqnarray}

Despite the above complicated form, we can straightforwardly see that using
the explicit critical points presented in Table \ref{crit}, the matrix
${\bf {Q}}$ acquires a simple form that allows for an easy calculation of
its eigenvalues. The corresponding eigenvalues $\nu_i$ ($i=1,2,3$)
for each critical point are presented in table
\ref{eigen}.

Thus, by determining the sign of the real parts of these
eigenvalues, we can classify the corresponding critical point. In
particular, if all the eigenvalues of a critical point have negative real
parts then the corresponding point is stable, if they all have positive
real parts then it is unstable, and if they change sign then it is a
saddle point.

\section{Stability of the critical points at infinity}\label{Poincare}

\begin{table*}[htp]
\begin{center}
\begin{tabular}{|c|c|c|c|}
\hline
 Cr. Point & Exists for & $\nu_1$ & $\nu_2$
\\
\hline \hline
$K_\pm$ &$\xi>0$,  all $\lambda$ & $\frac{1}{2}
\left(3-\sqrt{96 \xi -24
\sqrt{6\xi}  +9}\right)$ &
 $\frac{1}{2} \left(3+\sqrt{96 \xi -24 \sqrt{6\xi}  +9}\right)$\\
\hline $
L_\pm$ &   $\xi>0$,  all  $\lambda$  & $\frac{1}{2}
\left(3-\sqrt{96 \xi +24 \sqrt{6{\xi }
 }+9}\right)$ &
 $\frac{1}{2} \left(3+\sqrt{96 \xi +24 \sqrt{6\xi}  +9}\right)$ \\
   \hline
\end{tabular}
\end{center}
\caption[stability]{\label{eigen3}
 The eigenvalues of the matrix
${\bf {Q}}$ of the perturbation equations of the autonomous system
(\ref{Projection}), calculated for the four critical points at infinity.
Since we are restricted to the invariant set $y_r=0$,   ${\bf {Q}}$ is a
$2\times2$ matrix. }
\end{table*}

Let us consider the Poincar\'e variables
\begin{equation}
\label{Transf}
x_r=\rho \cos\theta \sin \psi ,\ z_r=\rho
\sin \theta \sin \psi ,\, y_r= \rho \cos \psi,\,
\end{equation} where $\rho=\frac{r}{\sqrt{1+r^2}},$ $r=\sqrt{x^2+y^2+z^2},$
$\theta\in[0,2\pi],$
and $-\frac{\pi}{2}\leq \psi \leq \frac{\pi}{2}$ (we
restrict the angle $\psi$ to this interval since the physical
region is given by $y>0$) \cite{PoincareProj,Leon:2008de,Leon:2010ai,Leon2011}.   Thus, the points at ``infinite''
($r\rightarrow+\infty$) are those having $\rho\rightarrow 1.$
Furthermore, the physical phase-space is given by
the intersection of the regions $2(x_r^2+y_r^2)\leq
1,\, x_r^2+y_r^2-z_r^2\geq 0$ and the circle
$x_r^2+y_r^2+z_r^2\leq 1$, that is it is the region
$$\left\{\left(x_r,y_r,z_r\right)\in [-1,1]\times [0,1]\times[-1,1]| z_r^2\leq x_r^2+y_r^2\leq
\frac{1}{2}\right\}.$$

Performing the transformation \eqref{Transf}, in terms of the Poincar\'e
variables the system \eqref{autonomous}  becomes:
\begin{align}
&x_r'=\frac{2  \sqrt{6\xi } \left(x_r^2-1\right) z_r x_r^2}{z_r^2+2
\left(x_r^2+y_r^2-1\right)}+\frac{3}{2} \left(2
   x_r^2-2 y_r^2-1\right) x_r +\frac{ \sqrt{6\xi } \left[\left(2
y_r^2-1\right) x_r^2-y_r^2+1\right]
   z_r}{x_r^2+y_r^2-1}\nonumber\\
&\ \ \ \ \ \ \ +\frac{\sqrt{\frac{3}{2}} \lambda
y_r^2}{\sqrt{-x_r^2-y_r^2-z_r^2+1}},\nonumber\\
&
 y_r'=
   \frac{2 \sqrt{6\xi } y_r z_r
   \left(-x_r^2-y_r^2-z_r^2+1\right) x_r^3}{\left(x_r^2+y_r^2-1\right)
\left[z_r^2+2
   \left(x_r^2+y_r^2-1\right)\right]}
+\frac{1}{2} y_r
   \left(6 x_r^2+4  \sqrt{6\xi } z_r x_r-6 y_r^2+3\right)\nonumber\\
&\ \ \ \ \ \ \
-\frac{\sqrt{\frac{3}{2}} \lambda  y_r
x_r}{\sqrt{-x_r^2-y_r^2-z_r^2+1}},\nonumber\\
& z_r'=
  -\frac{4  \sqrt{6\xi } \left(x_r^2+y_r^2-1\right)
   x_r^3}{z_r^2+2 \left(x_r^2+y_r^2-1\right)}+\frac{2  \sqrt{6\xi } y_r^2
z_r^2 x_r}{x_r^2+y_r^2-1}
+ \sqrt{6\xi }
   \left(2 x_r^3+x_r\right) \nonumber\\
&\ \ \ \ \ \ \
   +\frac{3 \left(2 x_r^4-x_r^2-2
y_r^4+y_r^2\right) z_r}{2 \left(x_r^2+y_r^2-1\right)}. \ \ \ \ \ \
\label{Projection}
\end{align}
Transforming to spherical coordinates and taking the limit $\rho\rightarrow
1,$ the leading terms in \eqref{Projection}
are
\begin{eqnarray}
&&\rho'\rightarrow 0,\label{inftya}\nonumber\\
&&\sqrt{1-\rho ^2} \theta'\rightarrow -\sqrt{\frac{3}{2}} \lambda {\cos\psi
\cot\psi \sin\theta},\label{inftyb}\nonumber\\
&& \sqrt{1-\rho ^2} \psi'\rightarrow \sqrt{\frac{3}{2}} \lambda  {\cos\theta \cos\psi}.\label{inftyc}
\end{eqnarray}
Since the equation for $\rho$ decouples, it is adequate to investigate the
subsystem of the angular variables.
Therefore, the singular points at infinity satisfy
\begin{eqnarray}
 &&x_r=\pm\cos\theta,\nonumber\\
&&z_r=\pm\sin\theta, \nonumber\\
 && y_r=0,\label{Pointsatinfinity}
\end{eqnarray}
for some particular
$\theta\in[0,2\pi]$ that are determined as follows.
Substituting \eqref{Pointsatinfinity} in \eqref{Projection} and solving for $\theta'$ we obtain
\begin{eqnarray}
\theta'=-\frac{1}{2} \cos (2 \theta ) \left(3 \cot \theta+2 \sqrt{6}
   \sqrt{\xi }\right).
\label{thetacoords}
\end{eqnarray}
This equation admits the parametric solution
\begin{eqnarray}
&&\tau (\theta )=c_1+\frac{1}{9-24 \xi }\left\{4 \sqrt{6\xi }
\tanh ^{-1}(\tan \theta )\right. \nonumber\\
&& \left.   \ \ \ \ \ \ \ \ \  +3 \ln [\cos (2
   \theta )]  -6 \ln \left[3 \cos \theta+2
\sqrt{6\xi } \sin
   \theta\right]\right\}.\nonumber
\end{eqnarray}
Thus, the singular solutions with $\theta\in[0,2\pi]$ are those satisfying
$\theta=\frac{\pi}{4},\frac{3\pi}{4},$ and
$\cot\theta=-2\sqrt{\frac{2}{3}}\xi$,  which due to
(\ref{Pointsatinfinity}) lead to simple expressions for $x_r$,$y_r$,$z_r$.
These results are summarized in Table \ref{crit2}.

In summary there are four physical critical points at infinity. The
points $K_\pm$, satisfying $ \theta=\frac{3\pi}{4}$ and the points $L_\pm$
satisfying $\theta=\frac{\pi}{4}$.  Thus, according to (\ref{Transf}),
these critical points correspond to the limit
$\phi\rightarrow \pm \infty, \dot\phi/H\rightarrow \pm \infty$ satisfying
the rate $$\left(\ln\phi\right)'\equiv
\sqrt{6}\frac{x}{z}=\sqrt{6}\cot\theta.$$

Concerning the stability analysis of the singular points  $K_\pm$  and
$L_\pm$, we take advantage that they are located in the invariant
set $y_r=0$ and we examine their stability for the reduced 2D
system $x_r,z_r$.  The eigenvalues of the linearized 2D
subsystem evaluated at $K_\pm$  and
$L_\pm$ are displayed in  Table
\ref{eigen3}, while the results of the stability analysis are presented in
Table  \ref{crit2}.

Finally, we mention that mathematically there are two more critical
points, namely the points $P_\pm$ satisfying
$\cot\theta=-2\sqrt{\frac{2}{3}}\xi$. However, they have no physical
meaning since they give a divergent $\Omega_{DE}$ in (\ref{Omegas22}),
namely $\text{sign}[8 \xi-3]\cdot\infty$, and thus we do not consider them
in the cosmological discussion.

\end{appendix}


\begin{thebibliography}{99}


\bibitem{Riess:1998cb}
  A.~G.~Riess {\it et al.} [Supernova Search Team Collaboration],
  {\it{Observational evidence from supernovae for an accelerating universe
and a cosmological constant}},
 Astron.\ J.\  {\bf 116}, 1009 (1998),
[\href{http://xxx.lanl.gov/abs/astro-ph/9805201}
{{\tt arXiv:astro-ph/9805201}}].

\bibitem{Perlmutter:1998np}
  S.~Perlmutter {\it et al.} [Supernova Cosmology Project Collaboration],
  {\it{Measurements of Omega and Lambda from 42 high redshift supernovae}},
  Astrophys.\ J.\  {\bf 517}, 565 (1999),
[\href{http://xxx.lanl.gov/abs/astro-ph/9812133}
{{\tt arXiv:astro-ph/9812133}}].



\bibitem{Bennett:2003bz}
  C.~L.~Bennett {\it et al.}  [WMAP Collaboration],
   {\it{First Year Wilkinson Microwave Anisotropy Probe (WMAP)
Observations: Preliminary Maps and Basic Results}},
  Astrophys.\ J.\ Suppl.\  {\bf 148}, 1 (2003),
[\href{http://xxx.lanl.gov/abs/astro-ph/0302207}
{{\tt arXiv:astro-ph/0302207}}].


 

\bibitem{Nojiri:2006ri}
  S.~Nojiri and S.~D.~Odintsov,
      {\it{Introduction to modified gravity and gravitational alternative
for dark
  energy}},
  eConf {\bf C0602061}, 06 (2006), Int.\ J.\ Geom.\ Meth.\ Mod.\ Phys.\ 
{\bf 4}, 115 (2007),
[\href{http://xxx.lanl.gov/abs/hep-th/0601213}
{{\tt arXiv:hep-th/0601213}}].

 

\bibitem{Ratra:1987rm}
  B.~Ratra and P.~J.~E.~Peebles,
     {\it{Cosmological Consequences of a Rolling Homogeneous Scalar
Field}},
  Phys.\ Rev.\  D {\bf 37}, 3406 (1988).

\bibitem{Wetterich:1987fm}
  C.~Wetterich,
     {\it{Cosmology and the Fate of Dilatation Symmetry}},
  Nucl.\ Phys.\  B {\bf 302}, 668 (1988).

\bibitem{Zlatev:1998tr}
  I.~Zlatev, L.~M.~Wang and P.~J.~Steinhardt,
     {\it{Quintessence, Cosmic Coincidence, and the Cosmological
Constant}},
  Phys.\ Rev.\ Lett.\  {\bf 82}, 896 (1999),
[\href{http://xxx.lanl.gov/abs/astro-ph/9807002}
{{\tt arXiv:astro-ph/9807002}}].

\bibitem{Boisseau:2000pr} 
  B.~Boisseau, G.~Esposito-Farese, D.~Polarski and A.~A.~Starobinsky,
     {\it{Reconstruction of a scalar tensor theory of gravity in an
accelerating universe}},
  Phys.\ Rev.\ Lett.\  {\bf 85}, 2236 (2000),
[\href{http://xxx.lanl.gov/abs/gr-qc/0001066}
{{\tt arXiv:gr-qc/0001066}}].

 


\bibitem{Guo:2006ab}
  Z.~K.~Guo, N.~Ohta and Y.~Z.~Zhang,
     {\it{Parametrizations of the dark energy density and scalar
potentials}},
  Mod.\ Phys.\ Lett.\  A {\bf 22}, 883 (2007),
[\href{http://xxx.lanl.gov/abs/astro-ph/0603109}
{{\tt arXiv:astro-ph/0603109}}].

\bibitem{Dutta:2009yb}
  S.~Dutta, E.~N.~Saridakis and R.~J.~Scherrer,
      {\it{Dark energy from a quintessence (phantom) field rolling near
potential
  minimum (maximum)}},
  Phys.\ Rev.\  D {\bf 79}, 103005 (2009),
[\href{http://xxx.lanl.gov/abs/0903.3412}
{{\tt arXiv:0903.3412}}].

 
\bibitem{Sahni:1998at}
  V.~Sahni and S.~Habib,
     {\it{Does inflationary particle production suggest Omega(m) < 1?}},
  Phys.\ Rev.\ Lett.\  {\bf 81}, 1766 (1998),
[\href{http://xxx.lanl.gov/abs/hep-ph/9808204}
{{\tt arXiv:hep-ph/9808204}}].

 

\bibitem{Uzan:1999ch}
  J.~P.~Uzan,
     {\it{Cosmological scaling solutions of non-minimally coupled scalar
fields}},
  Phys.\ Rev.\  D {\bf 59}, 123510 (1999),
[\href{http://xxx.lanl.gov/abs/gr-qc/9903004}
{{\tt arXiv:gr-qc/9903004}}].

 

\bibitem{Faraoni:2000wk}
  V.~Faraoni,
     {\it{Inflation and quintessence with nonminimal coupling}},
  Phys.\ Rev.\  D {\bf 62}, 023504 (2000),
[\href{http://xxx.lanl.gov/abs/gr-qc/0002091}
{{\tt arXiv:gr-qc/0002091}}].

  

\bibitem{Gong:2002dk}
  Y.~g.~Gong,
     {\it{Quintessence model and observational constraints}},
  Class.\ Quant.\ Grav.\  {\bf 19}, 4537 (2002),
[\href{http://xxx.lanl.gov/abs/gr-qc/0203007}
{{\tt arXiv:gr-qc/0203007}}].
 

\bibitem{Elizalde:2004mq}
  E.~Elizalde, S.~Nojiri and S.~D.~Odintsov,
      {\it{Late-time cosmology in (phantom) scalar-tensor theory: Dark
energy and  the
  cosmic speed-up}},
  Phys.\ Rev.\  D {\bf 70}, 043539 (2004),
[\href{http://xxx.lanl.gov/abs/hep-th/0405034}
{{\tt arXiv:hep-th/0405034}}].


 

\bibitem{Faraoni:2004dn}
  V.~Faraoni,
     {\it{de Sitter attractors in generalized gravity}},
  Phys.\ Rev.\  D {\bf 70}, 044037 (2004),
[\href{http://xxx.lanl.gov/abs/gr-qc/0407021}
{{\tt arXiv:gr-qc/0407021}}].


 

\bibitem{Setare:2008pc}
  M.~R.~Setare and E.~N.~Saridakis,
      {\it{Non-minimally coupled canonical, phantom and quintom models of
holographic
  dark energy}},
  Phys.\ Lett.\  B {\bf 671}, 331 (2009),
[\href{http://xxx.lanl.gov/abs/0810.0645}
{{\tt arXiv:0810.0645}}].
 
  
\bibitem{Caldwell2002a}
  R.~R.~Caldwell,
      {\it{A Phantom menace?}},
  Phys.\ Lett.\ B {\bf 545}, 23 (2002),
[\href{http://xxx.lanl.gov/abs/astro-ph/9908168}
{{\tt arXiv:astro-ph/9908168}}].

 

\bibitem{Caldwell:2003vq}
  R.~R.~Caldwell, M.~Kamionkowski and N.~N.~Weinberg,
     {\it{Phantom Energy and Cosmic Doomsday}},
  Phys.\ Rev.\ Lett.\  {\bf 91}, 071301 (2003),
[\href{http://xxx.lanl.gov/abs/astro-ph/0302506}
{{\tt arXiv:astro-ph/0302506}}].

 

\bibitem{Nojiri:2003vn}
  S.~Nojiri and S.~D.~Odintsov,
     {\it{Quantum deSitter cosmology and phantom matter}},
  Phys.\ Lett.\  B {\bf 562}, 147 (2003),
[\href{http://xxx.lanl.gov/abs/hep-th/0303117}
{{\tt arXiv:hep-th/0303117}}].


\bibitem{Onemli:2004mb}
  V.~K.~Onemli and R.~P.~Woodard,
     {\it{Quantum effects can render w $<$-1 on cosmological scales}},
  Phys.\ Rev.\  D {\bf 70}, 107301 (2004),
[\href{http://xxx.lanl.gov/abs/gr-qc/0406098}
{{\tt arXiv:gr-qc/04060987}}].


 

\bibitem{Saridakis:2009pj}
  E.~N.~Saridakis,
     {\it{Phantom evolution in power-law potentials}},
  Nucl.\ Phys.\  B {\bf 819}, 116 (2009),
[\href{http://xxx.lanl.gov/abs/0902.3978}
{{\tt arXiv:0902.3978}}].


 

\bibitem{Dutta:2009dr}
  S.~Dutta and R.~J.~Scherrer,
     {\it{Dark Energy from a Phantom Field Near a Local Potential
Minimum}},
  Phys.\ Lett.\  B {\bf 676}, 12 (2009),
[\href{http://xxx.lanl.gov/abs/0902.1004}
{{\tt arXiv:0902.1004}}].
 

\bibitem{Feng:2004ad}
  B.~Feng, X.~L.~Wang and X.~M.~Zhang,
     {\it{Dark Energy Constraints from the Cosmic Age and Supernova}},
  Phys.\ Lett.\  B {\bf 607}, 35 (2005),
[\href{http://xxx.lanl.gov/abs/astro-ph/0404224}
{{\tt arXiv:astro-ph/0404224}}].

 

\bibitem{Guo:2004fq}
  Z.~K.~Guo, Y.~S.~Piao, X.~M.~Zhang and Y.~Z.~Zhang,
     {\it{Cosmological evolution of a quintom model of dark energy}},
  Phys.\ Lett.\  B {\bf 608}, 177 (2005),
[\href{http://xxx.lanl.gov/abs/astro-ph/0410654}
{{\tt arXiv:astro-ph/0410654}}].
 
\bibitem{Feng:2004ff}
  B.~Feng, M.~Li, Y.~S.~Piao and X.~Zhang,
     {\it{Oscillating Quintom and the Recurrent Universe}},
  Phys.\ Lett.\  B {\bf 634}, 101 (2006),
[\href{http://xxx.lanl.gov/abs/astro-ph/0407432}
{{\tt arXiv:astro-ph/0407432}}].

 

\bibitem{Zhao:2006mp}
  W.~Zhao,
     {\it{Quintom models with an equation of state crossing -1}},
  Phys.\ Rev.\  D {\bf 73}, 123509 (2006),
[\href{http://xxx.lanl.gov/abs/astro-ph/0604460}
{{\tt arXiv:astro-ph/0604460}}].

 
\bibitem{Lazkoz:2006pa}
  R.~Lazkoz and G.~Leon,
     {\it{Quintom cosmologies admitting either tracking or phantom
attractors}},
  Phys.\ Lett.\  B {\bf 638}, 303 (2006),
[\href{http://xxx.lanl.gov/abs/astro-ph/0602590}
{{\tt arXiv:astro-ph/0602590}}].

 

\bibitem{Lazkoz:2007mx}
  R.~Lazkoz, G.~Leon and I.~Quiros,
     {\it{Quintom cosmologies with arbitrary potentials}},
  Phys.\ Lett.\  B {\bf 649}, 103 (2007),
[\href{http://xxx.lanl.gov/abs/astro-ph/0701353}
{{\tt arXiv:astro-ph/0701353}}].

 

\bibitem{Saridakis:2009jq}
  E.~N.~Saridakis and J.~M.~Weller,
     {\it{A Quintom scenario with mixed kinetic terms}},
  Phys.\ Rev.\  D {\bf 81}, 123523 (2010),
[\href{http://xxx.lanl.gov/abs/0912.5304}
{{\tt arXiv:0912.5304}}].

 

\bibitem{Setare:2008si}
  M.~R.~Setare and E.~N.~Saridakis,
     {\it{Quintom model with O($N$) symmetry}},
  JCAP {\bf 0809}, 026 (2008),
[\href{http://xxx.lanl.gov/abs/0809.0114}
{{\tt arXiv:0809.0114}}].

 

\bibitem{Cai:2009zp}
  Y.~F.~Cai, E.~N.~Saridakis, M.~R.~Setare and J.~Q.~Xia,
     {\it{Quintom Cosmology: Theoretical implications and observations}},
  Phys.\ Rept.\  {\bf 493}, 1 (2010),
[\href{http://xxx.lanl.gov/abs/0909.2776}
{{\tt arXiv:0909.2776}}].

 

\bibitem{Copeland:2006wr}
  E.~J.~Copeland, M.~Sami and S.~Tsujikawa,
     {\it{Dynamics of dark energy}},
  Int.\ J.\ Mod.\ Phys.\  D {\bf 15}, 1753 (2006),
[\href{http://xxx.lanl.gov/abs/hep-th/0603057}
{{\tt arXiv:hep-th/0603057}}].

 

\bibitem{Leon:2009ce}
  G.~Leon, Y.~Leyva, E.~N.~Saridakis, O.~Martin and R.~Cardenas,
     {\it{Falsifying Field-based Dark Energy Models}},
  in{\it{ Dark Energy: Theories, Developments, and Implications}}, Nova
Science Publishers,
(2010),
[\href{http://xxx.lanl.gov/abs/0912.0542}
{{\tt arXiv:0912.0542}}].

 

\bibitem{Geng:2011aj}
  C.~Q.~Geng, C.~C.~Lee, E.~N.~Saridakis and Y.~P.~Wu,
     {\it{'Teleparallel' Dark Energy}},
  Phys.\ Lett.\  B {\bf 704}, 384 (2011),
[\href{http://xxx.lanl.gov/abs/1109.1092}
{{\tt arXiv:1109.1092}}].
 

\bibitem{Geng:2011ka}
  C.~Q.~Geng, C.~C.~Lee and E.~N.~Saridakis,
     {\it{Observational Constraints on Teleparallel Dark Energy}},
  JCAP {\bf 1201}, 002 (2012),
[\href{http://xxx.lanl.gov/abs/1110.0913}
{{\tt arXiv:1110.0913}}].
 
\bibitem{Unzicker:2005in}
 A.~Unzicker and T.~Case,  
{\it{Translation of Einstein's attempt of a unified field
theory with teleparallelism}},
[\href{http://xxx.lanl.gov/abs/physics/0503046}
{{\tt arXiv:physics/0503046}}].

\bibitem{Hayashi:1979qx}
  K.~Hayashi and T.~Shirafuji,
     {\it{New general relativity}},
  Phys.\ Rev.\  D {\bf 19}, 3524 (1979)
  [Addendum-ibid.\  D {\bf 24}, 3312 (1982)].

\bibitem{Chen:2010va}
  S.~H.~Chen, J.~B.~Dent, S.~Dutta and E.~N.~Saridakis,
     {\it{Cosmological perturbations in f(T) gravity}},
  Phys.\ Rev.\  D {\bf 83}, 023508 (2011),
 [\href{http://xxx.lanl.gov/abs/1008.1250}
{{\tt arXiv:1008.1250}}].

\bibitem{Dent:2011zz}
  J.~B.~Dent, S.~Dutta and E.~N.~Saridakis,
      {\it{f(T) gravity mimicking dynamical dark energy. Background and
perturbation
  analysis}},
  JCAP {\bf 1101}, 009 (2011),
[\href{http://xxx.lanl.gov/abs/1010.2215}
{{\tt arXiv:1010.2215}}].

\bibitem{Cai:2011tc}
  Y.~F.~Cai, S.~H.~Chen, J.~B.~Dent, S.~Dutta and E.~N.~Saridakis,
     {\it{Matter Bounce Cosmology with the f(T) Gravity}},
  Class.\ Quant.\ Grav.\  {\bf 28}, 2150011 (2011),
[\href{http://xxx.lanl.gov/abs/1104.4349}
{{\tt arXiv:1104.4349}}].

 
  
\bibitem{Weitzenb23}
  Weitzenb\"{o}ck R.,
  \emph{Invarianten Theorie},
  (Nordhoff, Groningen, 1923).  

\bibitem{Maluf:1994ji}
  J.~W.~Maluf,
      {\it{Hamiltonian formulation of the teleparallel description of
general
  relativity}},
  J.\ Math.\ Phys.\  {\bf 35}, 335 (1994).

\bibitem{Arcos:2005ec}
  H.~I.~Arcos and J.~G.~Pereira,
     {\it{Torsion Gravity: a Reappraisal}},
  Int.\ J.\ Mod.\ Phys.\  D {\bf 13}, 2193 (2004),
[\href{http://xxx.lanl.gov/abs/gr-qc/0501017}
{{\tt arXiv:gr-qc/0501017}}].



 
\bibitem{Bengochea:2008gz}
  G.~R.~Bengochea and R.~Ferraro,
     {\it{Dark torsion as the cosmic speed-up}},
  Phys.\ Rev.\  D {\bf 79}, 124019 (2009),
[\href{http://xxx.lanl.gov/abs/0812.1205}
{{\tt arXiv:0812.1205}}].

 

\bibitem{Linder:2010py}
  E.~V.~Linder,
     {\it{Einstein's Other Gravity and the Acceleration of the Universe}},
  Phys.\ Rev.\  D {\bf 81}, 127301 (2010)
  [Erratum-ibid.\  D {\bf 82}, 109902 (2010)],
[\href{http://xxx.lanl.gov/abs/1005.3039}
{{\tt arXiv:1005.3039}}].


 

\bibitem{Myrzakulov:2010vz}
  R.~Myrzakulov,
     {\it{Accelerating universe from F(T) gravity}},
  Eur.\ Phys.\ J.\  C {\bf 71}, 1752 (2011),
[\href{http://xxx.lanl.gov/abs/1006.1120}
{{\tt arXiv:1006.1120}}].

 

\bibitem{Wu:2010av}
  P.~Wu and H.~W.~Yu,
     {\it{$f(T)$ models with phantom divide line crossing}},
  Eur.\ Phys.\ J.\  C {\bf 71}, 1552 (2011),
[\href{http://xxx.lanl.gov/abs/1008.3669}
{{\tt arXiv:1008.3669}}].


 

\bibitem{Bamba:2010iw}
  K.~Bamba, C.~Q.~Geng and C.~C.~Lee,
      {\it{Comment on 'Einstein's Other Gravity and the Acceleration of the
  Universe''}},
[\href{http://xxx.lanl.gov/abs/1008.4036}
{{\tt arXiv:1008.4036}}].

 

\bibitem{Zheng:2010am}
  R.~Zheng and Q.~G.~Huang,
     {\it{Growth factor in f(T) gravity}},
  JCAP {\bf 1103}, 002 (2011),
[\href{http://xxx.lanl.gov/abs/1010.3512}
{{\tt arXiv:1010.3512}}].

 

\bibitem{Bamba:2010wb}
  K.~Bamba, C.~Q.~Geng, C.~C.~Lee and L.~W.~Luo,
     {\it{Equation of state for dark energy in $f(T)$ gravity}},
  JCAP {\bf 1101}, 021 (2011),
[\href{http://xxx.lanl.gov/abs/1011.0508}
{{\tt arXiv:1011.0508}}].

 

\bibitem{Wang:2011xf}
  T.~Wang,
     {\it{Static Solutions with Spherical Symmetry in f(T) Theories}},
  Phys.\ Rev.\  D {\bf 84}, 024042 (2011),
[\href{http://xxx.lanl.gov/abs/1102.4410}
{{\tt arXiv:1102.4410}}].

 
\bibitem{Yerzhanov:2010vu}
  K.~K.~Yerzhanov, S.~R.~Myrzakul, I.~I.~Kulnazarov and R.~Myrzakulov,
     {\it{Accelerating cosmology in F(T) gravity with scalar field}},
[\href{http://xxx.lanl.gov/abs/1006.3879}
{{\tt arXiv:1006.3879}}].

  
\bibitem{Yang:2010ji}
  R.~J.~Yang,
     {\it{Conformal transformation in $f(T)$ theories}},
  Europhys.\ Lett.\  {\bf 93}, 60001 (2011),
[\href{http://xxx.lanl.gov/abs/1010.1376}
{{\tt arXiv:1010.1376}}].

 

\bibitem{Wu:2010mn}
  P.~Wu and H.~W.~Yu,
     {\it{Observational constraints on $f(T)$ theory}},
  Phys.\ Lett.\  B {\bf 693}, 415 (2010),
[\href{http://xxx.lanl.gov/abs/1006.0674}
{{\tt arXiv:1006.0674}}].

 

\bibitem{Bengochea:2010sg}
  G.~R.~Bengochea,
     {\it{Observational information for f(T) theories and Dark Torsion}},
  Phys.\ Lett.\  B {\bf 695}, 405 (2011),
[\href{http://xxx.lanl.gov/abs/1008.3188}
{{\tt arXiv:1008.3188}}].

 
\bibitem{Wu:2010xk}
  P.~Wu and H.~W.~Yu,
     {\it{The dynamical behavior of $f(T)$ theory}},
  Phys.\ Lett.\  B {\bf 692}, 176 (2010),
[\href{http://xxx.lanl.gov/abs/1007.2348}
{{\tt arXiv:1007.2348}}].

\bibitem{Li:2010cg}
  B.~Li, T.~P.~Sotiriou and J.~D.~Barrow,
     {\it{f(T) Gravity and local Lorentz invariance}},
  Phys.\ Rev.\  D {\bf 83}, 064035 (2011),
[\href{http://xxx.lanl.gov/abs/1010.1041}
{{\tt arXiv:1010.1041}}].

 

\bibitem{Zhang:2011qp}
  Y.~Zhang, H.~Li, Y.~Gong and Z.~H.~Zhu,
     {\it{Notes on $f(T)$ Theories}},
  JCAP {\bf 1107}, 015 (2011),
[\href{http://xxx.lanl.gov/abs/1103.0719}
{{\tt arXiv:1103.0719}}].

 

\bibitem{Deliduman:2011ga}
  C.~Deliduman and B.~Yapiskan,
     {\it{Absence of Relativistic Stars in f(T) Gravity}},
[\href{http://xxx.lanl.gov/abs/1103.2225}
{{\tt arXiv:1103.2225}}].

 

\bibitem{Chattopadhyay:2011fp}
  S.~Chattopadhyay and U.~Debnath,
     {\it{Emergent universe in chameleon, f(R) and f(T) gravity theories}},
  Int.\ J.\ Mod.\ Phys.\  D {\bf 20}, 1135 (2011),
[\href{http://xxx.lanl.gov/abs/1105.1091}
{{\tt arXiv:1105.1091}}].


\bibitem{Sharif:2011bi}
  M.~Sharif and S.~Rani,
     {\it{F(T) Models within Bianchi Type I Universe}},
  Mod.\ Phys.\ Lett.\  A {\bf 26}, 1657 (2011),
[\href{http://xxx.lanl.gov/abs/1105.6228}
{{\tt arXiv:1105.6228}}].


 

\bibitem{Li:2011rn}
  M.~Li, R.~X.~Miao and Y.~G.~Miao,
     {\it{Degrees of freedom of $f(T)$ gravity}},
  JHEP {\bf 1107}, 108 (2011),
[\href{http://xxx.lanl.gov/abs/1105.5934}
{{\tt arXiv:1105.5934}}].

 

\bibitem{Wei:2011jw}
  H.~Wei, X.~P.~Ma and H.~Y.~Qi,
     {\it{$f(T)$ Theories and Varying Fine Structure Constant}},
  Phys.\ Lett.\  B {\bf 703}, 74 (2011),
[\href{http://xxx.lanl.gov/abs/1106.01024}
{{\tt arXiv:1106.0102}}].

\bibitem{Ferraro:2011zb}
  R.~Ferraro and F.~Fiorini,
     {\it{Cosmological frames for theories with absolute parallelism}},
  Int.\ J.\ Mod.\ Phys.\ Conf.\ Ser.\  {\bf 3}, 227 (2011),
[\href{http://xxx.lanl.gov/abs/1106.6349}
{{\tt arXiv:1106.6349}}].

 

\bibitem{Miao:2011ki}
  R.~X.~Miao, M.~Li and Y.~G.~Miao,
      {\it{Violation of the first law of black hole thermodynamics in
$f(T)$
  gravity}},
  JCAP {\bf 1111}, 033 (2011),
[\href{http://xxx.lanl.gov/abs/1107.0515}
{{\tt arXiv:1107.0515}}].

 

\bibitem{Boehmer:2011gw}
  C.~G.~Boehmer, A.~Mussa and N.~Tamanini,
     {\it{Existence of relativistic stars in f(T) gravity}},
  Class.\ Quant.\ Grav.\  {\bf 28}, 245020 (2011),
[\href{http://xxx.lanl.gov/abs/1107.4455}
{{\tt arXiv:1107.4455}}].

 

\bibitem{Wei:2011mq}
  H.~Wei, H.~Y.~Qi and X.~P.~Ma,
     {\it{Constraining $f(T)$ Theories with the Varying Gravitational
Constant}},
[\href{http://xxx.lanl.gov/abs/1108.0859}
{{\tt arXiv:1108.0859}}].


 

\bibitem{Capozziello:2011hj}
  S.~Capozziello, V.~F.~Cardone, H.~Farajollahi and A.~Ravanpak,
     {\it{Cosmography in f(T)-gravity}},
  Phys.\ Rev.\  D {\bf 84}, 043527 (2011),
[\href{http://xxx.lanl.gov/abs/1108.2789}
{{\tt arXiv:1108.2789}}].

 

\bibitem{Wu:2011xa}
  P.~Wu and H.~Yu,
     {\it{The stability of the Einstein static state in $f(T)$ gravity}},
  Phys.\ Lett.\  B {\bf 703}, 223 (2011),
[\href{http://xxx.lanl.gov/abs/1108.5908}
{{\tt arXiv:1108.5908}}].
 

\bibitem{Daouda:2011rt}
  M.~H.~Daouda, M.~E.~Rodrigues and M.~J.~S.~Houndjo,
     {\it{Static Anisotropic Solutions in f(T) Theory}},
[\href{http://xxx.lanl.gov/abs/1109.0528}
{{\tt arXiv:1109.0528}}].


 

\bibitem{Bamba:2011pz}
  K.~Bamba and C.~Q.~Geng,
     {\it{Thermodynamics of cosmological horizons in $f(T)$ gravity}},
  JCAP {\bf 1111}, 008 (2011),
[\href{http://xxx.lanl.gov/abs/1109.1694}
{{\tt arXiv:1109.1694}}].


 

\bibitem{Wu:2011kh}
  Y.~P.~Wu and C.~Q.~Geng,
     {\it{Primordial Fluctuations within Teleparallelism}},
[\href{http://xxx.lanl.gov/abs/1110.3099}
{{\tt arXiv:1110.3099}}].


 

\bibitem{Gonzalez:2011dr}
  P.~A.~Gonzalez, E.~N.~Saridakis and Y.~Vasquez,
      {\it{Circularly symmetric solutions in three-dimensional
Teleparallel, f(T) and
  Maxwell-f(T) gravity}},
[\href{http://xxx.lanl.gov/abs/1110.4024}
{{\tt arXiv:1110.4024}}].


 

\bibitem{Ferraro:2011ks}
  R.~Ferraro and F.~Fiorini,
     {\it{Spherically symmetric static spacetimes in vacuum f(T) gravity}},
  Phys.\ Rev.\  D {\bf 84}, 083518 (2011),
[\href{http://xxx.lanl.gov/abs/1109.4209}
{{\tt arXiv:1109.4209}}].


\bibitem{Boehmer:2011si}
  C.~G.~Boehmer, T.~Harko and F.~S.~N.~Lobo,
      {\it{Wormhole geometries in modified teleparralel gravity and the
energy
  conditions}},
  Phys.\ Rev.\  D {\bf 85}, 044033 (2012),
[\href{http://xxx.lanl.gov/abs/1110.5756}
{{\tt arXiv:1110.5756}}].

 

\bibitem{Karami:2011nj}
  K.~Karami and A.~Abdolmaleki,
      {\it{Holographic and new agegraphic f(T)-gravity models with
power-law entropy
  correction}},
[\href{http://xxx.lanl.gov/abs/1111.7269}
{{\tt arXiv:1111.7269}}].


\bibitem{Wei:2011aa}
  H.~Wei, X.~J.~Guo and L.~F.~Wang,
     {\it{Noether Symmetry in $f(T)$ Theory}},
  Phys.\ Lett.\  B {\bf 707}, 298 (2012),
[\href{http://xxx.lanl.gov/abs/1112.2270}
{{\tt arXiv:1112.2270}}].


 

\bibitem{Atazadeh:2011aa}
  K.~Atazadeh and F.~Darabi,
     {\it{$f(T)$ cosmology via Noether symmetry}},
[\href{http://xxx.lanl.gov/abs/1112.2824}
{{\tt arXiv:1112.2824}}].


 

\bibitem{Farajollahi:2011af}
  H.~Farajollahi, A.~Ravanpak and P.~Wu,
     {\it{Cosmic acceleration and phantom crossing in $f(T)$-gravity}},
  Astrophys.\ Space Sci.\  {\bf 338}, 23 (2012),
[\href{http://xxx.lanl.gov/abs/1112.4700}
{{\tt arXiv:1112.4700}}].


 

\bibitem{Jamil:2011mc}
  M.~Jamil, D.~Momeni, N.~S.~Serikbayev and R.~Myrzakulov,
     {\it{FRW and Bianchi type I cosmology of f-essence}},
[\href{http://xxx.lanl.gov/abs/1112.4472}
{{\tt arXiv:1112.4472}}].

 

\bibitem{Karami:2012fu}
  K.~Karami and A.~Abdolmaleki,
     {\it{Generalized second law of thermodynamics in f(T)-gravity}},
[\href{http://xxx.lanl.gov/abs/1201.2511}
{{\tt arXiv:1201.2511}}].


 

\bibitem{Yang:2012hu}
  J.~Yang, Y.~L.~Li, Y.~Zhong and Y.~Li,
     {\it{Thick Brane Split Caused by Spacetime Torsion}},
[\href{http://xxx.lanl.gov/abs/1202.0129}
{{\tt arXiv:1202.0129}}].


 

\bibitem{Daouda:2012nj}
  M.~H.~Daouda, M.~E.~Rodrigues and M.~J.~S.~Houndjo,
     {\it{Anisotropic fluid for a set of non-diagonal tetrads in f(T)
gravity}},
[\href{http://xxx.lanl.gov/abs/1202.1147}
{{\tt arXiv:1202.1147}}].

\bibitem{Iorio:2012cm} 
  L.~Iorio and E.~N.~Saridakis,
     {\it{Solar system constraints on f(T) gravity}},
[\href{http://xxx.lanl.gov/abs/1203.5781}
{{\tt arXiv:1203.5781}}].
 
 

\bibitem{Copeland:1997et}
  E.~J.~Copeland, A.~R.~Liddle and D.~Wands,
     {\it{Exponential potentials and cosmological scaling solutions}},
  Phys.\ Rev.\  D {\bf 57}, 4686 (1998),
[\href{http://xxx.lanl.gov/abs/gr-qc/9711068}
{{\tt arXiv:gr-qc/9711068}}].

 

\bibitem{Ferreira:1997au}
  P.~G.~Ferreira and M.~Joyce,
     {\it{Structure formation with a self-tuning scalar field}},
  Phys.\ Rev.\ Lett.\  {\bf 79}, 4740 (1997),
[\href{http://xxx.lanl.gov/abs/astro-ph/9707286}
{{\tt arXiv:astro-ph/9707286}}].
 

\bibitem{Gong:2006sp}
  Y.~Gong, A.~Wang and Y.~Z.~Zhang,
     {\it{Exact scaling solutions and fixed points for general scalar
field}},
  Phys.\ Lett.\  B {\bf 636}, 286 (2006),
[\href{http://xxx.lanl.gov/abs/gr-qc/0603050}
{{\tt arXiv:gr-qc/0603050}}].
 

 

\bibitem{Chen:2008pz}
  X.~m.~Chen and Y.~Gong,
     {\it{Fixed points in interacting dark energy models}},
  Phys.\ Lett.\  B {\bf 675}, 9 (2009),
[\href{http://xxx.lanl.gov/abs/0811.1698}
{{\tt arXiv:0811.1698}}].

 

\bibitem{Chen:2008ft}
  X.~m.~Chen, Y.~g.~Gong and E.~N.~Saridakis,
     {\it{Phase-space analysis of interacting phantom cosmology}},
  JCAP {\bf 0904}, 001 (2009),
[\href{http://xxx.lanl.gov/abs/0812.1117}
{{\tt arXiv:0812.1117}}].

 
\bibitem{Schmidt:1990gb} 
  H.~-J.~Schmidt,
     {\it{New exact solutions for power law inflation Friedmann models}},
  Astron.\ Nachr.\  {\bf 311}, 165 (1990),
[\href{http://xxx.lanl.gov/abs/gr-qc/0109004}
{{\tt arXiv:gr-qc/0109004}}].

\bibitem{Muller:1989rp} 
  V.~Muller, H.~J.~Schmidt and A.~A.~Starobinsky,
     {\it{Power law inflation as an attractor solution for inhomogeneous
cosmological models}},
  Class.\ Quant.\ Grav.\  {\bf 7}, 1163 (1990).





\bibitem{Scherrer:2007pu}
  R.~J.~Scherrer and A.~A.~Sen,
     {\it{Thawing quintessence with a nearly flat potential}},
  Phys.\ Rev.\  D {\bf 77}, 083515 (2008),
[\href{http://xxx.lanl.gov/abs/0712.3450}
{{\tt arXiv:0712.3450}}].

 

\bibitem{Scherrer:2008be}
  R.~J.~Scherrer and A.~A.~Sen,
     {\it{Phantom Dark Energy Models with a Nearly Flat Potential}},
  Phys.\ Rev.\  D {\bf 78}, 067303 (2008),
[\href{http://xxx.lanl.gov/abs/0808.1880}
{{\tt arXiv:0808.1880}}].


\bibitem{Setare:2008sf}
  M.~R.~Setare and E.~N.~Saridakis,
     {\it{Quintom dark energy models with nearly flat potentials}},
  Phys.\ Rev.\  D {\bf 79}, 043005 (2009),
[\href{http://xxx.lanl.gov/abs/0810.4775}
{{\tt arXiv:0810.4775}}].


\bibitem{Wei:2011yr}
  H.~Wei,
     {\it{Dynamics of Teleparallel Dark Energy}},
[\href{http://xxx.lanl.gov/abs/1109.6107}
{{\tt arXiv:1109.6107}}].

 

\bibitem{PoincareProj} S., Lynch, {\it{Dynamical Systems with Applications
using Mathematica}}, Birkhauser,  Boston (2007).

\bibitem{Leon:2008de}
  G.~Leon,
     {\it{On the Past Asymptotic Dynamics of Non-minimally Coupled Dark
Energy}},
  Class.\ Quant.\ Grav.\  {\bf 26}, 035008 (2009),
[\href{http://xxx.lanl.gov/abs/0812.1013}
{{\tt arXiv:0812.1013}}].

 

\bibitem{Leon:2010ai}
  G.~Leon, P.~Silveira and C.~R.~Fadragas,
      {\it{Phase-space of flat Friedmann-Robertson-Walker models with both
a scalar
  field coupled to matter and radiation}},
    in {\it{ Classical and Quantum
Gravity: Theory, Analysis and Applications}}, Nova Science Publishers,
(2010),
[\href{http://xxx.lanl.gov/abs/1009.0689}
{{\tt arXiv:1009.0689}}].

\bibitem{Leon2011} G. Leon and C. R. Fadragas, {\it{Cosmological Dynamical Systems}}, LAP LAMBERT Academic Publishing,
(2011).


\bibitem{inprep}
  Y.~-F.~Cai, S.~-H.~Chen, J.~B.~Dent, S.~Dutta, E.~N.~Saridakis, in
preparation.




\end{thebibliography}
\end{document}